\documentclass[10pt,aps,prd,showpacs,twocolumn,superscriptaddress,longbibliography,nofootinbib,amsmath,amssymb,floatfix]{revtex4-1}
 \usepackage{amsmath}
 \usepackage{amssymb}
 \usepackage{multirow}
 \usepackage{subcaption}
 \usepackage{float}
\usepackage{graphicx}
 \usepackage{pstricks}
\usepackage{hyperref}

\begin{document}

\title{Stationary equilibrium torus supported by Weyssenhoff ideal spin fluid in Schwarzschild spacetime - I : Case of constant specific angular momentum distribution  }

\author{Sayantani Lahiri}
\email{sayantani.lahiri@zarm.uni-bremen.de}
\author{Claus Lämmerzahl}%
 \email{claus.laemmerzahl@zarm.uni-bremen.de}
 \affiliation{Center of Applied Space Technology and Microgravity (ZARM),\\
 28359 Bremen, Germany.}


\begin{abstract}
We consider a non-self-gravitating geometrically thick torus described by Weyssenhoff ideal spin fluid in a black hole spacetime. The Weyssenhof spin fluid shares the same symmetries of the background geometry,  i,e.  stationarity and axisymmetry and further describes circular orbital motion in the black hole spacetime. We further assume that assume the alignment of the spin is perpendicular to the equatorial plane. Under this set-up, we determine the integrability conditions of the general relativistic momentum conservation equation of Weyssenhoff ideal spin fluid using the Frenkel spin supplementary condition.
In the light of the integrability conditions, we then present stationary equilibrium solutions of the spin fluid torus with constant specific angular momentum distributions around the Schwarzschild black hole by numerically solving the general relativistic momentum conservation equation.  Our study reveals that both the iso-pressure and iso-density surfaces of torus get significantly modified in comparison to the ideal fluid torus without a spin fluid, owing to the spin tensor and its coupling to the curvature of the Schwarzschild black hole. In fact, the size of the torus is also found to be enhanced (diminished) depending on positive (negative) magnitude of spin parameter $s_0$. We finally estimate the magnitude of $s_0$ by assuming the torus to be composed of spin-1/2 particles.
\end{abstract}

\maketitle

\section{Introduction}
The existence of Black Holes (BH) is one of the most profound theoretical predictions of General Relativity. Generally gravitational collapse of massive stars are thought to be responsible for the existence of BHs \cite{PhysRev.56.455, PhysRevLett.14.57}.
Today, it is widely believed that almost all galaxies in the observable universe possess supermassive BH at their center \cite{Kormendy:2013dxa}, the significant confirmation of which comes out from the shadows of ultra-compact central objects \cite{Falcke:1999pj, Narayan:2013gca}. In recent times, an overwhelming support of this fact appeared from the images published by the Event Horizon Telescope Collaboration belonging to the shadow of the supermassive BH residing at the center of M87 galaxy \cite{EventHorizonTelescope:2019dse} and that of Sagittarius A$^{*}$ \cite{EventHorizonTelescope:2022xnr} at the center of Milky Way. 
The most direct confirmation is via the observation of the merger of BHs \cite{LIGOScientific:2016aoc}.
Additionally, there are also observational evidences for the existence of stellar-mass BHs through X-ray binaries \cite{1972Natur.235...37W, 1986ApJ...308..110M}. 

The accretion of matter onto BHs or any compact object causes efficient conversion of enormous gravitational energy of the infalling accreting matter into its rotational energy (in the Newtonian case) and to radiation, part of which contributes to the luminous disk-like structures \cite{2002apa..book.....F} surrounding the central compact object, commonly known as an accretion disk (AD).
   Generally, ADs are found in diverse astrophysical scenarios namely quasars, active galactic nuclei, X-ray binaries, the central engine of short GRBs and kilonovae \cite{Rees:1982pe, 1997PASJ...49..159F, Marscher2002ObservationalEF, Rezzolla:2010fd,Bildsten:1997vw,Popham:2000qa,1975Natur.256..628R}.  Depending on different features like the optical depth, mass accretion rate, geometrical thickness, they are categorised in different class of models namely geometrically thick disks or torus, thin disks, and advection-dominated accretion flows (ADAF) (see for example, \cite{AbramowiczFragile2013} and \cite{Rezzolla_book:2013}).
   
   The geometrical thick ADs, also known as Polish doughnuts, is the model which will consider hereafter.  These are stationary equilibrium configurations which can be constructed analytically.  Typically modeled by a relativistic hydrodynamical fluid, the equilibrium configurations were initially constructed using ideal fluid \cite{1976ApJ...207..962F, AbramowiczJaroszynskiSikora78, FoDai02, Lei:2008ui, KucSlStu11}. In recent years, different types of matter models have been considered for studying stationary equilibrium solutions, for e.g. with a viscous fluid \cite{ShakuraSunyaev73, Lahiri:2019mwc}, an electrically charged fluid \cite{Kovar11, KovarTr14,slany13}, or by a magnetized fluid \cite{Komissarov06, Gimeno-Soler:2017qmt, Lahiri:2020sza}.  The studies concerning properties and morphologies of ADs located in different BH geometries have also been addressed  in several works \cite{Rezzolla:2003re, Jefremov:2016dpi, Cruz-Osorio:2021gnz, Wielgus:2022qij, Bahamonde:2022jue, Cassing:2023bpt}. 
   The studies with accretion in presence of self-gravity effects can be found in many works \cite{1984MNRAS.208..279A,Lodato:2007hmf,Stergioulas:2011ga,Mach:2007zz,Mach:2022wrk}.

It is also likely that the constituents of an torus possess intrinsic spin angular momentum whose cumulative effects induce an overall non-zero spin of the fluid. 
Although spin is a microscopic property of matter, for the macroscopic description of matter, a corresponding hydrodynamical theory can be obtained by an averaging technique from the microscopic theory of matter with spin \cite{1960PThPh..24..291H}. 
A prominent example is the phenomenological Weyssenhoff fluid model which describes classical ideal hydrodynamical fluid with spin \cite{Weyssenhoff:1947iua} where the fluid elements are characterized by an intrinsic angular momentum, i,e. {\it spin} proportional to the volume. In this model, the spin angular momentum density is described by a second-rank anti-symmetric tensor $S^{\mu \nu}$ whose spatial components is a three-vector that coincides with three-density of the spinning matter in the rest frame. 
Motivated by the attempt to look for experiments/observations for torsion, the Lagrangian formulation of the Weyssenhoff spin fluid model has been developed in the context of Einstein-Cartan theory by taking torsion into consideration \cite{Obukhov:1987yu}.
Moreover, using the Lagrangian formulation, it has been shown by Obukhov and Piskareva \cite{ObukhovPiskareva1989} that in General Relativity, under the pole-dipole approximation, the conservation laws of the Weyssenhoff ideal spin fluid follows the generalized version of the evolution equation of spinning \textit{test} particles given by Mathisson \cite{Mathisson:1937zz}, Papapetrou \cite{1951RSPSA.209..248P} and Dixon \cite{1970RSPSA.314..499D}, commonly known as Mathisson-Papapetrou-Dixon (MPD) equations.
Additionally, in order to close the MPD equations a spin supplementary condition (SSC) has to be stated. The SSC is not unique. 
Several SSCs have been prescribed in the literature \cite{1956AcPP...15..389P,Frenkel:1926zz,Tulczyjew,Dixon:1970zza,newton1949localized,Kyrian:2007zz}. 
Notably, the conservation laws of the Weyessenhoff ideal spin fluid are obtained by adopting the Frenkel SSC \cite{Obukhov:1987yu, ObukhovPiskareva1989}.
Analogous to test particles with spin, the spin of the Weyssenhoff fluid interacts with the curvature of the background geometry leading to an additional spin-curvature coupling contribution in the momentum balance. 

In the present work, we examine the consequences of spin-curvature coupling on stationary equilibrium solutions of relativistic torus modeled by the Weyssenhoff ideal spin fluid. 
We have adopted the test-fluid approximation, thus, neglected any self-gravity effects of the torus.   
We start with a stationary and axisymmetric BH background where the spin fluid describes circular orbits and is endowed with constant specific orbital angular momentum distribution.
In this background geometry, the integrability conditions of the general relativistic momentum conservation equation of ideal Weyssenhoff spin fluid are derived by considering the spin being aligned perpendicular to the equatorial plane. 
Our study reveals the newly obtained integrability conditions of the spin fluid embodies the integrability conditions of ideal fluid without spin i.e. relativistic Von-Zeipel conditions. Additional conditions are obtained due to presence of spin tensor and its coupling to the spacetime curvature of the stationary, axisymmetric BH geometry. 

In order to determine the impacts of spin and curvature on stationary solutions of a torus, we consider the simplest non-rotating background geometry, i,e. a Schwarzschild BH.
Using the integrability condition, we then present the allowed structure of the spin tensor necessary to produce macroscopic spin effects on the morphology of the torus.
Here, we have consider the scenario when the fluid within the torus exactly fills its Roche Lobe.
Under this situation, the equilibrium stationary solutions of the torus are constructed by semi-analytically solving the general relativistic momentum conservation equation of Weyssenhoff fluid with constant specific angular momentum distributions. The corresponding solutions are then used to construct the iso-pressure and the iso-density surfaces of the torus.
By comparing with a torus described with only ideal fluid without spin, our analysis reveals that
the spin tensor-curvature coupling plays crucial role in modifying the the overall morphology of the torus as well as altering the locations of its cusp, center and outer edge.
The qualitative effects of the spin on the torus is characterised by a parameter $s_0$ and the size and total energy density of the torus are found to be either enhanced or reduced depending on the magnitude and sign of $s_0$.
In a simplest scenario, we have provided an estimation of $s_0$ by assuming that the torus is built with spin-1/2 particles.

The present paper is organised as follows. The Section II begins with the mathematical framework of the Weyssenhoff ideal spin fluid model. Following this, the integrability conditions of the general relativistic momentum conservation equation of Weyssenhoff spin fluid are presented. In Section III, we determine the complete structure of the spin tensor of the Weyssenhoff spin fluid which undergoes circular motion in the Schwarzschild spacetime. 
Using this, the stationary solutions of the equilibrium torus are constructed in the Schwarzschild BH spacetime. We discuss the results pertaining to the effects of spin and curvature on the morphology of the torus by studying the iso-density and iso-pressure surfaces constructed from the stationary solutions in Section IV. Finally, after giving an estimation for $s_0$, in Section V, a summary of our study is presented. \\
\textit{Notations and conventions}: We take Riemannian geometry with the signature $-,+,+,+$. The covariant derivative is denoted by $\nabla_\mu$. We will use geometrized units ($G = c = 1$) throughout the paper. The Greek indices represent the co-ordinate basis which runs from $t,r,\theta,\phi$.

\section{Physical framework}

\subsection{The matter model}
The aim of our work is to construct equilibrium solutions of a stationary torus composed of a fluid made of neutral particles with spin which can be of classical or quantum origin.
In the continuum limit such kind of matter can be described by the Weyssenhoff ideal neutral spin fluid \cite{Weyssenhoff:1947iua}.
In GR, the symmetric energy momentum tensor of the Weyssenhoff spin fluid model is given by \cite{ObukhovPiskareva1989}, 
\begin{equation}
T_{\mu\nu} = (\epsilon+p) u_\mu u_\nu + p g_{\mu \nu} + 2(g^{\rho\sigma}-u^{\rho}u^{\sigma}) \nabla_{\rho} [u_{(\mu}S_{\nu)\sigma}] \, , 
    \label{EM1}
\end{equation}
where $\epsilon$ and $p$ are the energy density and pressure of the fluid. The round brackets in \eqref{EM1} denote the symmetrization of $\mu$ and $\nu$ indices.  The 4-velocity $u^\mu$ of the spinning fluid constituents is normalized as $u^\mu u_\mu = -1$. The projection tensor to the particle's rest frame is given by $\triangle^{\mu\nu} = g^{\mu\nu} + u^\mu u^\nu$. Finally, $S^{\mu\nu}$ is the antisymmetric spin tensor that corresponds to the dipole contribution in the context of multipole moment expansion \cite{Weyssenhoff:1947iua}. 
The Weyssenhoff spin fluid model employs the Frenkel SSC, alternatively known as Mathisson-Pirani SSC, given by,
\begin{equation}
S^{\mu\nu} u_{\nu} =0 \, .  \label{Frenkel}
\end{equation}
This defines $u^\mu$ a the 4-velocity of the center of mass (CM) of the spinning body. 

The divergence of \eqref{EM1} and using the SSC \eqref{Frenkel} gives the momentum balance 
\begin{equation}
(\epsilon + p) a_\mu + \partial_\mu p + 2 \nabla_\rho (u^\rho S_{\mu\sigma} a^\sigma) + R_{\rho\sigma\tau\mu} S^{\rho\sigma} u^\tau = 0 \label{E1}
\end{equation}
where the Riemann curvature tensor is defined as
\begin{equation}
R^\mu{}_{\nu\rho\delta} = \partial_\rho \Gamma^\mu_{\delta\nu} - \partial_\delta \Gamma^\mu_{\rho\nu} + \Gamma^\lambda_{\delta\nu} \Gamma^\mu_{\rho\lambda} - \Gamma^\lambda_{\rho\nu} \Gamma^\mu_{\delta\lambda}
\end{equation}
and $\Gamma^\mu_{\rho\sigma}$ are the Christoffel symbols. 
The energy balance equation reads as follows
\begin{equation}
D\epsilon+(\epsilon+p) \nabla_\mu u^\mu = 0 \, , \label{E2}
\end{equation}
where $D = u^\nu \nabla_\nu$ and the 4-acceleration is given by $a^\mu = u^\nu \nabla_\nu u^{\mu}$. Eq. (\ref{E1}) represents the MPD equation in our spin fluid model. We note that in absence of spin (\ref{E1}) reduces to the Euler equation.

The spin four-vector is defined as,
\begin{equation}
    S_{\mu} = -\frac12 \epsilon_{\mu\nu\rho \sigma} u^{\nu} S^{\rho\sigma} \label{sv}
\end{equation}
and the inverse relation is given by
\begin{equation}
    S^{\mu\nu} = -\epsilon^{\mu\nu\rho \sigma} S_{\rho}u_{\sigma} \label{StensorfromSvector}
\end{equation}
The spin density scalar is defined by
\begin{equation}
   S^2= \frac12 S_{\mu\nu} S^{\mu\nu} \, . \label{SpinScalarDensty}
\end{equation}
$S$ depends on $r$ and $\theta$ but is constant along each particle trajectory. Let us define the following quantity,
\begin{equation}
    S^{\mu}{}_{\rho\sigma} = u^\mu S_{\rho\sigma},
\end{equation}
and by taking the divergence, one obtains,
\begin{eqnarray}
  \nabla_\mu (u^\mu S^{\rho\sigma}) &=& u^\rho u_\lambda \nabla_\mu(u^\mu S^{\lambda \sigma}) - u^\sigma u_\lambda \nabla_\mu(u^\mu S^{\lambda\rho})  \nonumber \\
  &=& (u^\sigma S^{\lambda\rho} - u^\rho S^{\lambda\sigma}) a_\lambda  \label{spin-eqn}
\end{eqnarray}
which implies the divergence of $S^{\mu}{}_{\rho\sigma}$ is vanishing if the four-acceleration also vanishes or is orthogonal to the spin tensor.

\subsection{Symmetries}

In order to find solutions of Weyssenhoff ideal spin fluid equation of motion we need to assume following symmetry conditions:
\begin{itemize}
\item[(i)] The BH background space-time is stationary and axisymmetric. The corresponding Killing vectors are given by $\eta^\mu = \partial_t = (1,0,0,0)$ and $\xi^\mu = \partial_\phi = (0,0,0,1)$. Accordingly, the Lie derivative along $\xi$ and $\eta$ of all geometric quantities vanish.  

\item[(ii)] The Weyssenhoff spin fluid shares the same symmetries of the background geometry, therefore any (tensorial) flow parameter $f$, including the spin, satisfies the conditions ${\cal L}_\eta f = 0$ and ${\cal L}_\xi f = 0$, where ${\cal L}$ is the Lie-derivative. Accordingly, in an adapted coordinate system all quantities depend on $r$ and $\theta$ only. This represents a major restriction. However, our aim is to determine the order of magnitude the spin induced changes of the shape of ADs for which special situations are fine. 

\item[(iii)] The spin fluid describes circular orbits and therefore the four-velocity of the fluid is given by 
\begin{equation}
u^\mu = (u^t,0,0,u^\phi) \, .
\end{equation}
This fluid does not constitute a stationary congruence. The angular velocity then is
\begin{equation}
\Omega = \frac{d\phi}{dt} = \frac{u^\phi}{u^t}
\end{equation}
and the specific orbital angular momentum is
\begin{equation}
- l = \frac{p_\phi}{p_t} = \frac{g_{\phi \mu} u^\mu}{g_{t\nu} u^\nu} = \frac{g_{\phi t} u^t + g_{\phi\phi} u^\phi}{g_{tt} u^t + g_{t\phi} u^\phi} \, . 
\end{equation}
We have
\begin{equation}
l = - \frac{g_{\phi t} + g_{\phi\phi} \Omega}{g_{tt} + g_{t\phi} \Omega}  \, , \qquad  \Omega = - \frac{g_{t\phi} + g_{tt} l}{g_{\phi\phi} + g_{t\phi} l} \, .
\end{equation}
The 4-velocity can be written as
\begin{equation}
u^\mu = A (\eta^\mu + \Omega \xi^\mu) \label{4vel}
\end{equation}
with
\begin{eqnarray}
A = u^t & = & \frac{1}{\sqrt{g_{tt} + 2 g_{t\phi} \Omega + g_{\phi\phi} \Omega^2}} \\ 
-u_{t} & = & \displaystyle\sqrt{\frac{g_{t\phi}^2-g_{tt}g_{\phi \phi}}{l^2g_{tt}+2lg_{t \phi}+g_{\phi \phi}}} \, .
\end{eqnarray}
The four-acceleration then can be computed as
\begin{equation}
  a_\mu = \partial_\mu \ln u_t -\displaystyle\frac{\Omega \partial_\mu l}{(1-\Omega l)} = \partial_\mu \ln u^t - \frac{l \partial_\mu \Omega}{1 - \Omega l} . \label{4accel}
\end{equation}

Such smooth orbital conditions are not compatible will all SSCs \cite{Kyrian:2007zz}.

\item[(iv)] The spin four-vector $S^\nu$ is aligned perpendicular to the equatorial plane, i.e. the spin vector is polar and is given by,
\begin{equation}
    S^\nu = S^\theta \delta^\nu_\theta \, .  \label{spin:1}
\end{equation}
Then the non-zero components of the spin tensor can be calculated from (\ref{StensorfromSvector})
\begin{eqnarray}
    S^{tr} & = & - S\sqrt{\frac{g_{\theta\theta}}{-g}} u_\phi  \label{compspin1} \\
  S^{r\phi} & = & -S\sqrt{\frac{g_{\theta \theta}}{-g}}u_{t} \label{compspin2}
\end{eqnarray}
where $S$ is the spin scalar density \eqref{SpinScalarDensty} and is given by $S= \sqrt{g_{\theta \theta}}S^{\theta}$. 
\end{itemize}
\subsection{Integrability conditions}
 In this section, we will determine the existence of integrability conditions of \eqref{E1}. Using Eq.(\ref{spin-eqn}), one obtains,
\begin{eqnarray}
   \nabla_\rho (u^\rho S_{\mu\sigma} a^\sigma)
   &=& g_{\mu \alpha}g_{\sigma \beta} \nabla_{\rho}(u^{\rho}S^{\alpha \beta}a^{\sigma})\nonumber \\
  &=& g_{\mu \alpha}g_{\sigma \beta}\left[\nabla_{\rho}(u^{\rho}S^{\alpha \beta})a^{\sigma}+u^{\rho}S^{\alpha \beta}(\nabla_{\rho}a^{\sigma})\right] \nonumber \\
  &=& g_{\mu \alpha}g_{\sigma \beta}\left[(u^{\beta}S^{\lambda \alpha}-u^{\alpha}S^{\lambda \beta})a_{\lambda}\right]+S_{\mu \sigma}Da^{\sigma} \nonumber \\
  &=& S^{\lambda}_{\;\; \mu}a_{\lambda}(u_{\sigma}a^{\sigma})-S^{\lambda \beta}a_{\lambda}a_{\beta}u_{\mu} +S_{\mu \sigma}Da^{\sigma} \nonumber \\
   &=& S_{\mu \beta} Da^{\beta}  
\end{eqnarray} 
where $u_{\sigma}a^{\sigma}=0$ and $S^{\lambda \beta}a_{\lambda}a_{\beta}=0$. Hence \eqref{E1} reduces to,
\begin{equation}
  (\epsilon+p) a_\mu + \partial_\mu p +2S_{\mu \beta} Da^{\beta} + R_{\rho\sigma\tau\mu} S^{\rho\sigma} u^\tau =0.  \label{mom-eqn}
\end{equation}
Note that \eqref{mom-eqn} further reduces to the Euler equation in absence of the spin.
Since the fluid undergoes circular motion, the rate of the change in acceleration can be considered to be proportional to the 4-acceleration. This implies that one can make a following ansatz for the term  $ Da^{\beta}$ in \eqref{E1} as follows,
\begin{equation}
    Da^{\beta} = \, \Omega^{ \beta \nu} a_{\nu}  \label{omega-tensor}
\end{equation}
where $\Omega_{\alpha \beta}$ is the antisymmetric angular velocity tensor satisfying $\Omega_{\alpha \beta}=-\Omega_{\beta \alpha}$.
Using \eqref{4vel}, the last term in \eqref{E1} can be written as,
\begin{eqnarray}
R_{\rho\sigma\tau\mu} S^{\rho\sigma} u^\mu  = A \left(R_{\rho\sigma\tau\mu} S^{\rho\sigma} \eta^\mu + \Omega R_{\rho\sigma\tau\mu} S^{\rho\sigma} \xi^\mu\right) \, . 
\end{eqnarray}
Let us make an ansatz and express it as follows,
\begin{eqnarray}
  R_{\alpha \beta \gamma \mu} S^{\alpha \beta}\eta^{\gamma}=  \partial_{\mu} \Phi, \label{eq:10}\\
 R_{\alpha \beta \gamma \mu} S^{\alpha \beta}\xi^{\gamma}=  \partial_{\mu} \phi \label{eq:11}
\end{eqnarray}
where the scalars $\phi$ and $\Phi$ are functions of $r$ and $\theta$. 
Note that ${\cal L}_{\eta}(R_{\alpha \beta \gamma \mu} S^{\alpha \beta}) = {\cal L}_{\xi}(R_{\alpha \beta \gamma \mu} S^{\alpha \beta})=0$ since due to the symmetries, the Lie derivatives of Riemann curvature tensor and spin tensor along the Killing vectors $\eta$ and $\xi$ vanish.  
Substituting \eqref{omega-tensor}, \eqref{eq:10}, \eqref{eq:11}, and \eqref{4accel} into \eqref{mom-eqn} we obtain,
\begin{eqnarray}
\frac{\partial_\mu p}{\epsilon+p} &=& -\partial_\mu\ln({-u_t}) + \frac{\Omega \partial_\mu l}{1 - \Omega l} \nonumber \\
&-& \frac{2S_{\mu \alpha}\Omega^{\alpha \beta}a_{\beta}}{\epsilon+p} -\frac{A \left(\partial_\mu \Phi + \Omega \partial_\mu \phi\right)}{\epsilon+p}  
\label{rad-pol} 
\end{eqnarray}
or, equivalently, 
\begin{eqnarray}
\frac{\partial_\mu p}{\epsilon+p} &=& -\partial_\mu\ln({-u^t}) + \frac{l \partial_\mu\Omega}{1 - \Omega l}  \nonumber \\
&-& \frac{S_{\mu \alpha}\Omega^{\alpha \beta}a_{\beta}}{\epsilon+p} -\frac{A \left(\partial_\mu \Phi + \Omega \partial_\mu \phi\right)}{\epsilon+p} \, .  \label{rad-polb} 
\end{eqnarray}

The integrability condition then reads
\begin{eqnarray}
0 & = & \frac{\partial_{[\mu} p \partial_{\nu]} \epsilon}{(\epsilon + p)^2} - \frac{\partial_{[\mu} l \partial_{\nu]} \Omega}{(1 - \Omega l)^2} \nonumber\\
& & - \partial_{[\mu} \Phi \partial_{\nu]} \left(\frac{A}{\epsilon + p}\right) + \partial_{[\mu} \phi \partial_{\nu]} \left(\frac{\Omega A}{\epsilon + p}\right) \,  \nonumber\\
& & - \partial_{\mu}\left( \frac{2 S_{\nu \alpha}\Omega^{\alpha \beta}a_{\beta}}{\epsilon+p}\right)+\partial_{\nu}\left( \frac{2S_{\mu \alpha}\Omega^{\alpha \beta}a_{\beta}}{\epsilon+p}\right), \label{integrability} 
\end{eqnarray}  \label{int-calc}
where the square brackets denote antisymmetrization. Here all terms are independent from the others: the first term is on energy and pressure, the second on angular velocity and momentum, and the last two are on spin coupling. Accordingly, we require that each term should vanish separately. The first term vanishes provided $\partial_\mu \epsilon \sim \partial_\mu p$, the second term vanishes for $\partial_\mu\Omega \sim \partial_\mu l$ and for the other two terms we require $\partial_\mu \Phi \sim \partial_\mu\frac{A}{\epsilon + p}$ and $\partial_\mu \phi \sim \partial_\mu\frac{\Omega A}{\epsilon + p}$. Since the last two terms of \eqref{int-calc} must also vanish, this then means
\begin{eqnarray}
\epsilon & = & \epsilon(p) \label{ic-1} \\ 
\Omega & = & \Omega(l) \label{ic-2} \\
\Phi & = & \Phi\left(\frac{A}{\epsilon + p}\right) \label{ic-3} \\
\phi & = & \phi\left(\frac{\Omega A}{\epsilon + p}\right)  \label{ic-4} \, \\
\partial_{\mu} B&=& \frac{2 S_{\mu \alpha}\Omega^{\alpha \beta}a_{\beta}}{\epsilon+p}  \label{B-form},
\end{eqnarray}
that is, we obtain a number of equations of states between $\epsilon$ and $p$ (barotropic equation of state), between $\Omega$ and $l$, between $\Phi$ and $\frac{A}{\epsilon + p}$, and between $\phi$ and between $\frac{\Omega A}{\epsilon + p}$.  
The integrability condition of an ideal fluid in absence of spin is only given by from \eqref{ic-1}\footnote{The integrability condition of the Euler equation in absence of spin is commonly known as the Von-Zeipel condition. \cite{AbramowiczJaroszynskiSikora78}.}.
 In the stationary and axisymmetric spacetime, where the spin of the fluid undergoing circular motion is aligned along a specific orientation (see, \eqref{spin:1}), one readily concludes that the function $B$ is independent of the co-ordinates $t$ and $\phi$.
Therefore from \eqref{B-form}, one obtains for $\mu=t$ and  $\mu=\phi$ the following conditions.
\begin{eqnarray}
    \partial_{t} B &=& S_{t r}\Omega^{rr}a_{r} + S_{t r}\Omega^{r \theta}a_{\theta}=0 \\
    \partial_{\phi} B &=& S_{\phi r}\Omega^{r r}a_{r} + S_{\phi r}\Omega^{\phi \theta}a_{\theta}=0
\end{eqnarray}
Given the fact that $\Omega^{rr}=0$, it follows that $\Omega^{r \theta}=0 $ from the above conditions.
Also $ \partial_{\theta} B= S_{\theta \alpha} \Omega^{\alpha r}a_{r} + S_{\theta \alpha} \Omega^{\alpha \theta}a_{\theta}
    = 0$
as $S_{\theta \alpha}=0$ due to \eqref{spin:1}. Therefore the integrability conditions holds provided $B=B(r)$ and is determined from \eqref{B-form} using \eqref{compspin1} and \eqref{compspin2}. 
Hence \eqref{mom-eqn} is expressed in the integral form as,
\begin{eqnarray}
\ln |u_{t}| &-&\int_{0}^{l} \frac{\Omega dl}{1-\Omega l} +\int_{0}^{p}\frac{dp}{\epsilon+p} +\int d B\nonumber \\
&+&\int_{0}^{\Phi}\frac{A(r,\theta)d\Phi}{\epsilon+p}+
\int_{0}^{\phi}\frac{\Omega A(r,\theta)d\phi}{\epsilon+p} =\text{const}
\nonumber \\
\label{Potential}. 
\end{eqnarray}
At the surface and on the inner edge of the disk, both the pressure and the energy density of the fluid vanish.  Also at the radial position of inner edge i,e. $r=r_{\mathrm{in}}$, one can usually express $l=l_{in}$, $u_{t}= u_{t_{in}}$ and $\Phi_{in}=0=\phi_{in}$. Therefore Eq.(\ref{Potential}) becomes,
\begin{eqnarray}
W&-&W_{\mathrm{in}}+\int_{0}^{p}\frac{dp}{\epsilon+p} \\ \nonumber
&+&\int_{0}^{r_{\mathrm{in}}} d B+\int_{0}^{\Phi}\frac{A(r,\theta)d\Phi}{\epsilon+p}+
\int_{0}^{\phi}\frac{\Omega A(r,\theta)d\phi}{\epsilon+p}=0
\end{eqnarray}
The total potential is given by,
\begin{eqnarray}
W-W_{\mathrm{in}} = \ln{|u_{t}|}- \ln{|u_{t_{\mathrm{in}}}|}- \int_{l_{\mathrm{in}}}^{l} \frac{\Omega dl}{1-\Omega l}
\end{eqnarray} 
and $W_{\mathrm{in}}$ is the total potential at the inner edge of the torus at the equatorial plane.
\section{Equilibrium tori with Weyssenhoff spin fluid}
\label{sec-soln-tori}

\subsection{Methodology}
Let us now consider a relativistic non-self gravitating torus described by the Weyssenhoff ideal spin fluid in static, non-rotating BH spacetime, described by the Schwarzschild metric. 
In a simplified set-up, the torus is characterized by constant specific orbital angular momentum distribution. The spin fluid in the torus is in hydrostatic equilibrium, shares the same symmetries of the Schwarschild BH (see, \eqref{spin:1}-\eqref{compspin2}) and undergoes purely circular orbits.   
In addition, we assume the internal energy density, $\varepsilon$, is very small and, therefore, the total energy is approximately equal to the rest-mass density i,e. $\epsilon=\rho(1+\varepsilon) \approx \rho$.
In Schawarzschild co-ordinates, the Schwarzschild metric is given by,
\begin{equation}
    ds^2=g_{tt}dt^2 +g_{rr}dr^2+ g_{\theta \theta}d \theta ^2+ g_{\phi \phi} d \phi ^2
\end{equation}
where $0 <t <\infty $, $0\leq r\leq \infty$, $0\leq \theta \leq \pi $ and $0\leq \phi \leq 2 \pi$. The metric coefficients are,
\begin{eqnarray}
g_{tt}=-\left(1-\displaystyle\frac{2M}{r}\right)= -\frac{1}{g_{rr}},  \quad  
g_{\theta \theta}=r^2,   \, 
g_{\phi \phi}= r^2 \sin^2\theta   \nonumber \\
\end{eqnarray} 
where $M$ is mass parameter of the Schwarzschild BH which is set as $M=1$ in the rest of the paper.
As $g_{t \phi} =0$ for a non-rotating spcatime, the specific angular momentum and the angular velocity are related as,
\begin{eqnarray}
\frac{l(r, \theta)}{\Omega(r, \theta)}=-\frac{g_{\phi \phi}}{g_{tt}}
\end{eqnarray}
With the assumption of constant specific angular momentum distributions, we set $l(r, \theta)=l_0$ where $l_0$ is a constant and in accordance to the integrability condition (see \eqref{ic-1}), we consider the equation of state as $p=\kappa \epsilon^{\gamma}$, so that \eqref{mom-eqn} is expressed as,
\begin{eqnarray}
\partial_{\mu} \epsilon =\frac{(\epsilon^{2-\gamma}+\kappa \epsilon)}{-\kappa \gamma}\partial_{\mu} \ln(-u_t{})- \frac{(2S_{\alpha \beta}Da^{\beta}+R_{\rho \sigma \tau \mu}S^{\rho \sigma}u^{\tau})}{\kappa \gamma \epsilon^{\gamma-1}} \label{mom-eqn1}\nonumber \\ 
\end{eqnarray}
where $\kappa$ is the polytropic constant and $\gamma$ is the adiabatic coefficient. Using \eqref{compspin1} and \eqref{compspin2}, radial and polar components of \eqref{mom-eqn1} can be read off as follows,
\begin{eqnarray}
  \partial_{r} \epsilon &=& \frac{(\epsilon^{2-\gamma}+\kappa \epsilon)}{-\kappa \gamma}\partial_{r}\ln (-u_{t}) -\frac{S(r,\theta)}{\kappa \gamma\epsilon^{\gamma-1}(1-\Omega l_0)}\sqrt{\frac{g_{\theta \theta}}{-g}} F_1\nonumber \\
&-& \frac{2g_{rr}g_{tt}S(r,\theta)}{\kappa \gamma\epsilon^{\gamma-1}} \sqrt{\frac{g_{\theta \theta}}{-g}} \,[u_{\phi} Da^{t}-\frac{g_{\phi \phi}}{g_{tt}}u_{t}Da^{\phi}]
\label{den-1}\nonumber \\
 \\[2mm]
\partial_{\theta}\epsilon &=& \frac{(\epsilon^{2-\gamma}+\kappa \epsilon)}{-\kappa \gamma}\partial_{\theta}\ln (-u_{t})-\frac{S(r,\theta)}{\kappa \gamma\epsilon^{\gamma-1}(1-\Omega l_0)}\sqrt{\frac{g_{\theta \theta}}{-g}}F_2
\nonumber \\ 
\label{den-2} 
\end{eqnarray}
where $F_1$ and $F_2$ are given by,
\begin{eqnarray}
F_1(r,\theta)&=& - \Omega l_0 R_{tr\phi r}+\Omega R_{r \phi \phi r}- l_0 R_{trtr}+R_{r \phi t r}\,, \nonumber \\
F_2(r,\theta)&=& - \Omega l_0 R_{tr\phi \theta }+\Omega R_{r \phi \phi \theta}- l_0 R_{tr t\theta}+ R_{r \phi t \theta} \nonumber 
\end{eqnarray}
To obtain $F_1$ and $F_2$ following relations are used,
\begin{eqnarray}
u_{\phi}u^{t} &=& \frac{l}{1-\Omega l_0},\quad 
u_{t}u^{\phi}=-\frac{\Omega}{1-\Omega l_0}. \label{relations}
\end{eqnarray}
The non-zero components of the curvature tensor appearing in $F_1(r,\theta)$ and $F_2(r,\theta)$ are,
\begin{eqnarray}
R_{trtr}=-\frac{2}{r^3}, \qquad  R_{r\phi \phi r}=\frac{\sin^2\theta}{(r-2)}
\end{eqnarray}
so that $F_1(r,\theta)$ and $F_2(r,\theta)$ finally reduces to,  
\begin{eqnarray}
 F_1(r,\theta) = \frac{3l_0}{r^3}, \quad F_2(r,\theta) =0. \label{eq:19}
\end{eqnarray}
The stationary solutions of the equilibrium ideal spin fluid torus are then obtained by computing the energy density and the pressure which are determined by solving \eqref{den-1} and \eqref{den-2} for a spin length function $S$ along with the values of the constants $\kappa$ and $\gamma$. 

The existence of integrability conditions imply 
$dp/(\epsilon+p)$ is an exact differential (see, \eqref{Potential}), as a result, the compatibility condition $\partial_{\theta} \partial_{r}p=\partial_{r} \partial_{\theta}p$ expressed as the commutation of partial second derivatives of the fluid pressure is automatically satisfied. 
We utilize the compatibility condition for determining the spin length function as discussed in the following section.
\\

\subsection{Determination of spin length function}
In order to obtain the solutions of fluid pressure and energy density, we must assign values for $\kappa, \gamma$,  and determine the spin length function S. In view of the fact that the Weyssenhoff fluid shares the same symmetries as that of the background geometry, $S$ is therefore purely a function of radial and polar co-ordinates.
The value of the adiabatic index is chosen to be $\gamma =2$.   The fluid pressure and the energy density are then related by the equation of state as $p= \kappa \epsilon ^2$. Note that this choice is high for any physical systems, but taking this value of the adiabatic index makes it convenient to perform analytic integration for our simple model of spin fluid torus presented here.
We start with a general form of $S$ as,
\begin{equation}
 S(r,\theta) = s_0 k(r,\theta)\epsilon^{\gamma-1}(1-\Omega l_0) \label{eq:17}
\end{equation}
Here $s_0$ is a constant that can take both positive and negative values. \\
Substituting $\gamma=2$ in \eqref{eq:17}, \eqref{den-1} and \eqref{den-2}, one immediately obtains,
\begin{eqnarray}
      \partial_{r} \epsilon & =& -\frac{(1+\kappa \epsilon)}{2\kappa}\partial_{r}\ln (-u_{t}) \nonumber \\ [2mm]
    &-&\frac{l_0 s_0 k \csc \theta \left[l_0^2 (r-2)^2 \csc ^2\theta \left(3-r \csc ^2\theta \right)+(r-3) r^3\right]}{\kappa  (r-2) r^4 \left(r^3-l_0^2
   (r-2) \csc ^2\theta\right)} \nonumber \\
   \label{eq:15}\\[2mm]
\partial_{\theta}\epsilon & =& -\frac{(1+\kappa \epsilon)}{2\kappa}\partial_{\theta}\ln (-u_{t}). \label{eq:16}
\end{eqnarray}
Note that \eqref{eq:15} and \eqref{eq:16} reduce to the Euler equation valid for an ideal fluid for $s_0=0$,. 
With our choice of the equation of state, the compatibility condition on the fluid pressure also implies second derivatives of energy density commute as $\partial_{\theta} \partial_r \epsilon=\partial_{r} \partial_{\theta} \epsilon $.
Substituting \eqref{eq:15}, \eqref{eq:16} in the compatibility condition, the first order differential equation for $k(r,\theta)$ can be read off as,
 \begin{eqnarray}
     \partial_{\theta}k+\frac{P_1(r,\theta)}{P_2(r,\theta)}k=0 \label{eq-k}
 \end{eqnarray}
 where,
 \begin{widetext}
 \begin{eqnarray}
     \frac{P_1(r,\theta)}{P_2(r,\theta)}=\frac{\cot \theta \left[l_0^2 (r-2)  \left \{(r-2) \csc ^2\theta \left(20 r^4-10 l_0^2 (r-2) r \csc ^2\theta+3 l_0^2 (r-2)\right)+3 r^3 (24-11 r) \right \}+2r^6 \sin ^2 \theta (12-5 r) \right]}{2\, l_0^2 (r-2) \left[(r-2) \csc ^2\theta \left(2\, l_0^2 (r-2) r
   \csc ^2 \theta -3 l_0^2 (r-2)-2 r^4\right)-2 r^3 (r-3) \right]-2 r^6 \sin ^2\theta(12-5 r) } \nonumber
 \end{eqnarray}
 \end{widetext}
The solution of $k(r,\theta)$ is obtained as follows by solving \eqref{eq-k} analytically,
\begin{equation}
    k(r,\theta) = \frac{ 2^{1/4} \sin ^{5/2}\theta \left[l_0^2 (r-2)-r^3 \sin ^2\theta \right]^{5/4}C_1(r)}{4 l_0^2 (r-2)^2 \left(2 r-3 \sin
   ^2\theta \right)+4 (12-5 r) r^3 \sin ^4\theta} \label{soln-k}
\end{equation}
where $C_1(r)$ is a constant of integration over $\theta$ and is fixed at the equatorial plane. Following \eqref{soln-k}, $C_1(r)$ at the equatorial plane can be readily expressed in terms of $k(r,\pi/2)$ and $l_0$ in the following way,
\begin{equation}
C_1(r)=\displaystyle k\left(r,\frac{\pi}{2}\right) \frac{4 l_0^2 (r-2)^2 \left(2 r-3 \right)+4 (12-5 r) r^3}{2^{1/4} \left[l_0^2 (r-2)-r^3 \right]^{5/4}}
\end{equation} 
For convenience and simplicity of the computations, we choose $k(r, \pi/2)=1$ which is the simplest possible choice at par with the integrability conditions.
Using \eqref{soln-k} and \eqref{eq:17}, the spin length function in terms of the energy density, $s_0$ and $l_0$ is finally given by, 
\begin{widetext}
\begin{equation}
 S(r, \theta)= s_0   \frac{\sin ^{5/2}\theta  \left[l_0^2 \,(r-2)^2 (2 r-3)+ r^3(12-5 r) \right]}{\left[ l_0^2\, (r-2)^2 \left(2 r-3 \sin ^2\theta \right)+ r^3(12-5 r) \sin ^4 \theta\right]} \left[\frac{l_0^2 \,(r-2)-r^3 \sin ^2\theta }{l_0^2 (r-2)-r^3} \right]^{5/4} \epsilon(r,\theta) (1-\Omega l_0)\label{sf}
\end{equation}
\end{widetext}
With the determination of the solutions of $\epsilon(r,\theta)$, the spin length function can be easily computed using \eqref{sf} for the values of $s_0$ and $l_0$.
\subsection{Results}
We now obtain the stationary solutions of the energy density and pressure of the equilibrium torus modelled by Weyssenhoff ideal spin fluid. By integrating Eq. (\ref{eq:16}) directly, the expression for the energy density is found to be,
\begin{equation}
    \epsilon(r, \theta) = \frac{2^{1/4}\kappa C_2(r)(r^3 \sin^2 \theta-l_0^2(r-2))^{1/4}-\sqrt{\sin \theta}}{\kappa \sqrt{\sin \theta}} \label{energy1}
\end{equation}
and $C_2(r)$ is the radial co-ordinate dependent constant of integration independent of $\theta$. It is determined from the first order ordinary differential equation at the equatorial plane (about which the spin is defined by \eqref{spin:1}) by substituting \eqref{energy1} in  \eqref{eq:15}  as follows,
\begin{eqnarray}
\frac{d C_2}{dr}&+&\frac{(3 r-4)}{4r(r-2)} C_2 \nonumber \\
&-&\frac{l_0 s_0 \left[l_0^2 (r-2)^2 (2 r-3)+r^3(12-5 r) \right]}{2 \kappa. \, 2^{1/4}
    (r-2) r^4 \left[r^3-l_0^2 (r-2)\right]^{5/4}} =0 \nonumber \\
    \label{C2-eqn1}
\end{eqnarray}
The above equation is solved numerically under appropriate boundary conditions which when substituted in \eqref{energy1} generates the iso-density and iso-pressure surfaces of the torus. We next describe the scheme for obtaining $C_2(r)$.

To demonstrate the effects of spin fluid on equilibrium structure of a torus, we mainly concentrate on closed torus configuration which is characterized by its cusp $r_{\mathrm{cusp}}$ that corresponds to the radial location where the iso-pressure and iso-density surfaces self-cross, center $r_{\mathrm{c}}$ that corresponds to location of maximum pressure and energy density, the inner edge $r_{\rm{in}}$ and an outer edge $r_{\mathrm{out}}$, all defined at the equatorial plane.
Furthermore, we build stationary models of equilibrium torus for which the fluid exactly fills the Roche Lobe. 
In terms of the effective potential, this amounts to the potential gap $\triangle W$ reducing to  $\triangle W = W_{\rm{in}}- W_{\rm{cusp}} =0$ where  $W_{\rm{in}}$ and $W_{\rm{cusp}}$ are the respective effective potentials of the inner edge and the cusp at the equatorial plane \cite{RezZanFon03}. Consequently, the inner edge of the torus coincides with the cusp at the equatorial plane. 

Since for an exactly filling torus configuration with $\triangle W=0$ corresponds to $p(r_{\mathrm{in}})=\epsilon(r_{\mathrm{in}})=0$ at $r_{\rm{in}}=r_{\mathrm{cusp}}$,
using the condition $\epsilon(r_{\rm {in}})=0$ one then obtains the boundary condition for solving
the differential equation for \eqref{C2-eqn1} as follows,
\begin{equation}
    C_2(r_{\rm{in}}) = \frac{1}{2^{1/4}\kappa (r_{\rm{in} }^3 -l_0^2(r_{\rm{in}}-2))^{1/4}} \label{C2-1}
\end{equation}
The inner edge of the torus must be known to compute $ C_2(r_{\rm{in}})$ which serves as the boundary condition for numerically solving \eqref{C2-eqn1} for given values of $l_0$ and $s_0$.  
Note that the cusp of the torus corresponds to the location of the extrema where isopressure and isodensity surfaces self-cross. 
In the present context, it coincides with the inner edge, therefore imposing the condition 
 $\partial_{r} \epsilon(r=r_{\rm {in}})=0$ in \eqref{eq:15}, together with \eqref{energy1} and \eqref{C2-1}, the polynomial equation for determining the inner edge of the torus is found to be,
 \begin{eqnarray}
 r_{\mathrm{in}}^6-l_0^2 (r_{\mathrm{in}}-2)^2 r_{\mathrm{in}}^3&-&l_0^3 (r_{\mathrm{in}}-2)^2 (2 r_{\mathrm{in}}-3)s_0 \nonumber \\
& +&l_0 s_0(5 r_{\mathrm{in}}-12) \,r_{\mathrm{in}}^3 =0 \label{rin-eqn}
\end{eqnarray}
where we have further set $\kappa =1$. 
Out of all six possible roots of \eqref{rin-eqn}, the real positive root of $r_{\rm in}>r_{\mathrm{Sch}}$ is chosen, where $r_{\mathrm{Sch}}$ corresponds to the Schwarzschild radius.\\
On a different note, it can be easily observed that for $s_0=0$, one retrieves the Keplarian distribution given by $l_0=\displaystyle\frac{r\sqrt{r}}{r-2}=l_{\mathrm{k}}$, where $l_{\mathrm{k}}$ is the Keplarian specific angular momentum at $r_{\mathrm{in}}$ coinciding $r_{\mathrm{cusp}}$ \cite{FoDai02}.

\begin{figure*}[htb!]
\captionsetup{justification=raggedright, singlelinecheck=off}
\centering
\begin{subfigure}[b]{\columnwidth}
\captionsetup{justification=centering}
        \centering
 \includegraphics[scale=0.43]{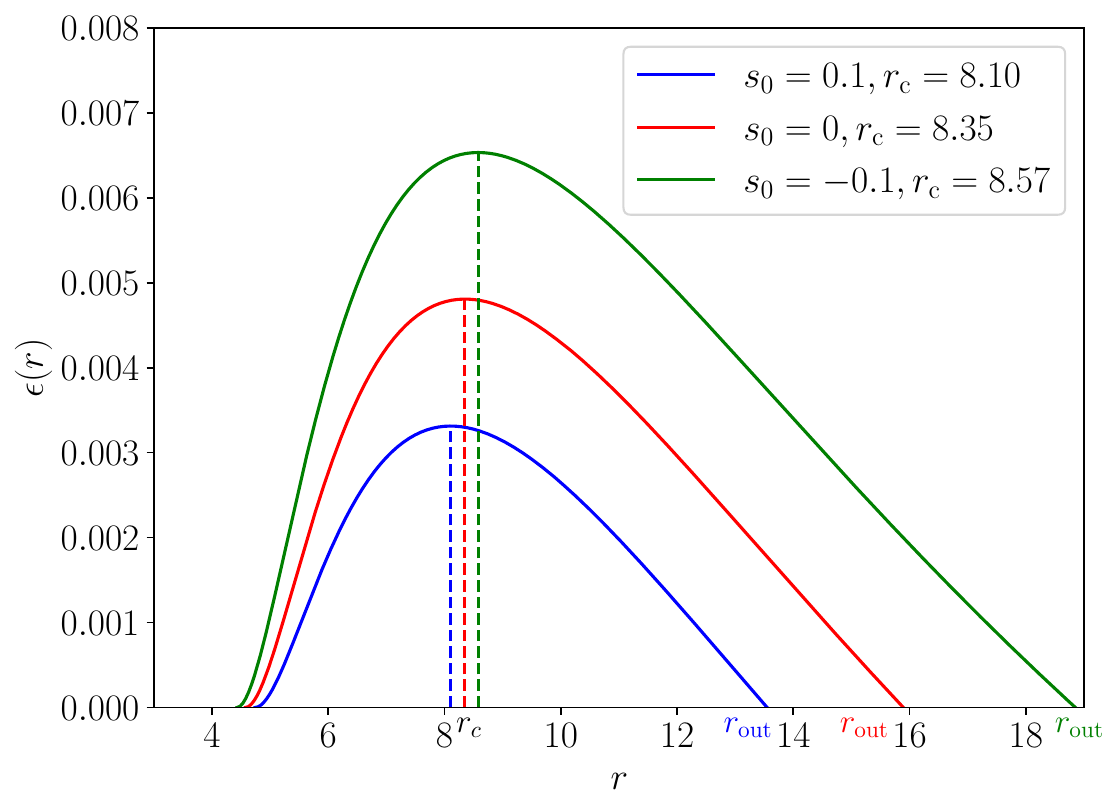} 
 \caption{} \label{fig1a}
  \end{subfigure}%
 \hfill
    \begin{subfigure}[b]{\columnwidth}
     \captionsetup{justification=centering}
        \centering
  \includegraphics[scale=0.43]{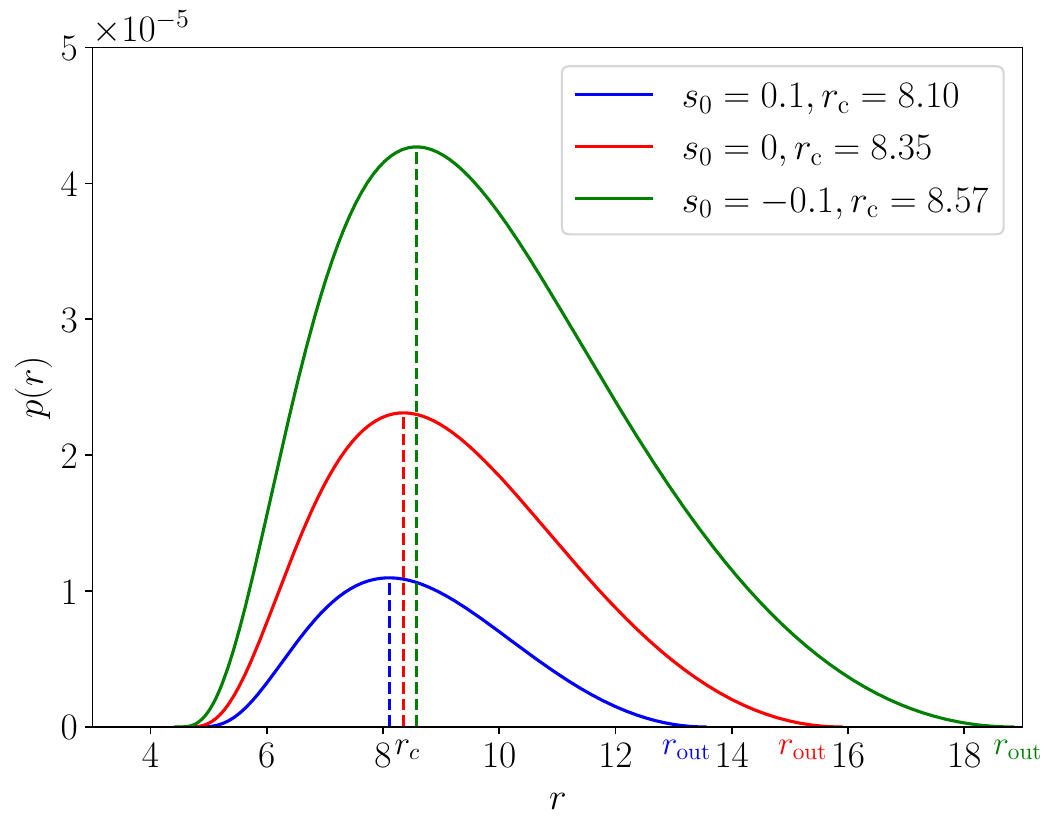}   
  \caption{} \label{fig1b}
\end{subfigure}%
\caption{ Iso-density and iso-pressure profiles of Weyssenhoff spin torus at the equatorial plane for $s_0>0, s_0=0$ and $s_0<0$ with $l_0=3.8$. The red dotted line is the radial location of the center $r_{\rm{c}}$ of a torus without spin fluid ($s_0= 0$). The shift in the radial position of $r_c$ occurs by changing the magnitude of $s_0$ as shown by the blue dotted line for $s_0 = 0.1$ and the green dotted line for $s_0 = -\, 0.1$.  For each $s_0$, the outer edge of the torus $r_{\mathrm{out}}$ is shown. }
\end{figure*}

The iso-density surfaces of the equilibrium torus are constructed with two constant values of specific angular momentum, namely  $l_0=3.8$ and $l_0= 4$.
We first consider $l_0=3.8$, for this value of $l_0$,  solutions of closed equilibrium tori are obtained characterized by cusp, center and an well-defined outer edge.
The constant energy density profiles of such closed torus at the equatorial plane are shown
in fig. \ref{fig1a} for $s_0=0$, $s_0 =0.1$ and $s_0 =-0.1$.
Let us now discuss the procedure for obtaining the solution of the energy density with $s_0=0.1$. First, $r_{\mathrm{in}}$ is evaluated by solving \eqref{rin-eqn} which gives rise to two positive real roots, namely, $r_{\mathrm{in}}=4.7424$ and $r_{\mathrm{in}}=8.1051$. 
Substituting $r_{\mathrm{in}}$ in \eqref{C2-1} and solving \eqref{C2-eqn1} numerically with $C_2(r_\mathrm{in})$ produces the solution of the energy density function \eqref{energy1}. A closed torus structure possessing a cusp, center and an outer edge only exists for $r_{\mathrm{in}}=4.7424$ whereas with $r_{\mathrm{in}}=8.1051$, no torus structure could be found. Similarly, for $s_0=-0.1$ we choose $r_{\mathrm{in}}$ that generates a torus configuration.

The impacts of spin and its coupling to the spacetime curvature of the BH geometry on equilibrium solutions can be appreciated from both the iso-energy density and iso-pressure contours, as shown respectively in fig. \ref{fig1a} and fig. \ref{fig1b}.
Clearly, the radial location of the newly formed center of torus in presence of spin never coincides with the torus center filled with ideal fluid without spin contributions. 
Both fig. \ref{fig1a} and fig. \ref{fig1b} show this feature where that the center $r_{c}$ of the spin fluid torus gets slightly shifted in comparison to a purely ideal fluid torus. 
This can be observed for both $s_0>0$ and  $s_0<0$.
When compared to the torus without spin fluid ($s_0=0$), in the former case, the location of $r_{c}$ moves towards the horizon and the corresponding magnitude of the maximum energy density well as pressure at $r_{c}$ get lowered whereas in the latter case, $r_{c}$ moves away from the horizon. 
The maximum energy density as well as the maximum pressure at $r_{c}$ get further enhanced for $s_0<0$ in comparison to the torus characterized by $s_0=0$.  
Additionally, the size the torus also gets diminished in the $s_0>0$ parameter region.  On the other hand, it increases for all values with $s_0<0$ and can be clearly noticed from the changes in positions of $r_{\rm{in}}$ and $r_{\rm{out}}$ when compared to the equilibrium torus without Weyssenhoff ideal fluid.

\begin{figure}[htb!]
\captionsetup{justification=raggedright, singlelinecheck=off}
\centering
\begin{subfigure}[b]{\columnwidth}
\captionsetup{justification=centering}
        \centering
 \includegraphics[scale=0.43]{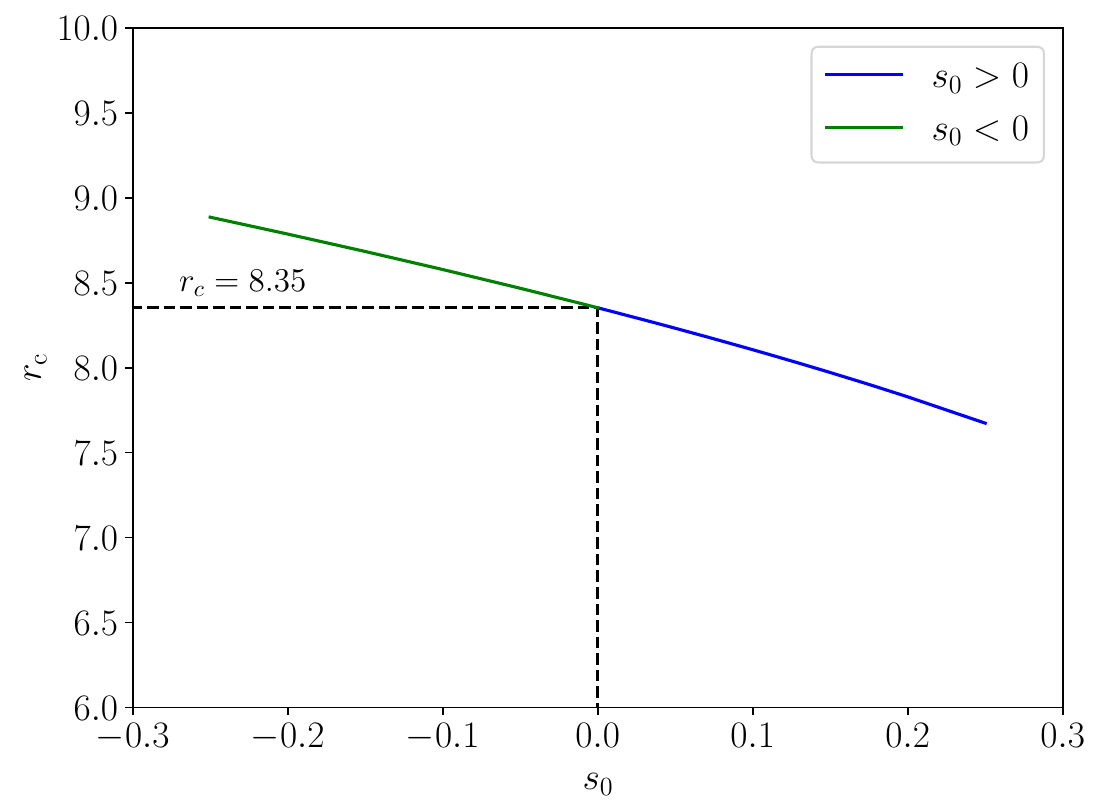} 
 \caption{} \label{fig2a}
  \end{subfigure}%
 \hfill
    \begin{subfigure}[b]{\columnwidth}
     \captionsetup{justification=centering}
        \centering
  \includegraphics[scale=0.43]{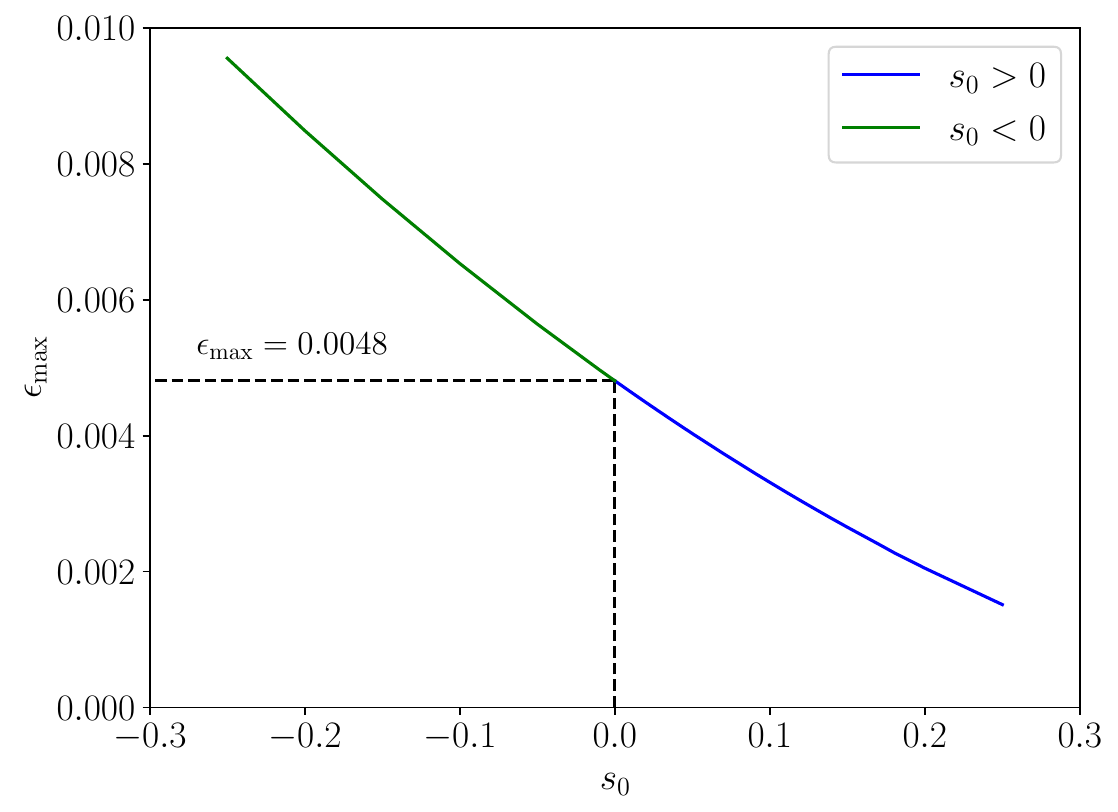}   
  \caption{} \label{fig2b}
\end{subfigure}%
 \caption{Behaviour of $r_c$ and $\epsilon_{\mathrm{max}}$ with changing $s_0$ at the equatorial plane. The black dotted lines in the upper and lower panels denote respective values of $\epsilon_{\mathrm{max}}$ and $r_c$ for an ideal fluid torus without spin contributions.  }
\end{figure}

The effect of changing $s_0$ on the radial position of the center $r_c$ and the corresponding 
energy density (denoted by $\epsilon_{\rm{max}}$) can be noticed from fig. \ref{fig2a} and fig. \ref{fig2b} respectively. When compared with a purely ideal fluid torus ($s_0=0$), both $r_c$ and $\epsilon_{\rm{max}}$ monotonically decrease as $s_0$ is increased from $s_0 =0$ whereas if the alignment of the spin is reversed about the equatorial plane, i.e. in the $s_0<0$ region, the radial position of $r_c$ and $\epsilon_{\rm{max}}$ increase monotonically. 
In essence, $r_c$ moves away from the horizon for all $s_0<0$ values, resulting into an enhancement of $\epsilon_{\rm{max}}$. On the other hand, $r_c$ comes closer to the BH horizon by increasing $s_0$ with positive values leading to a lowering of $\epsilon_{\mathrm{max}}$ in comparison to a torus without the spin fluid.

\begin{figure*}[htb!]
\captionsetup{justification=raggedright, singlelinecheck=on}
\hspace{-1.55cm}
  \includegraphics[scale=0.52]{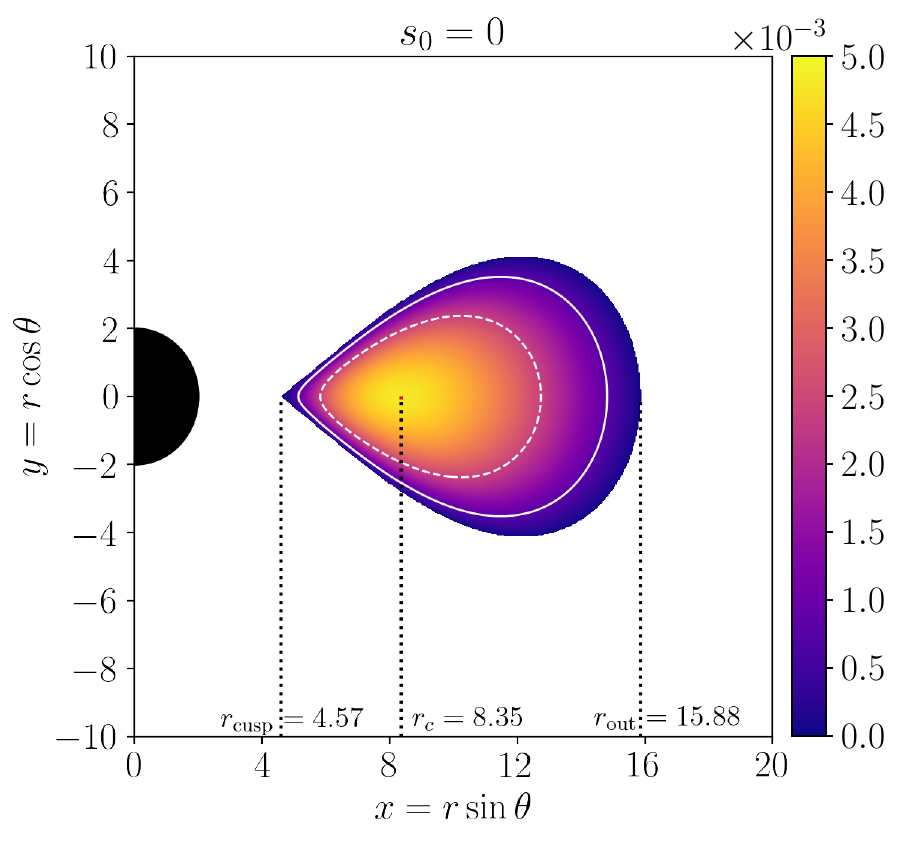}
  	\hspace{-0.1cm}
  	\includegraphics[scale=0.52]{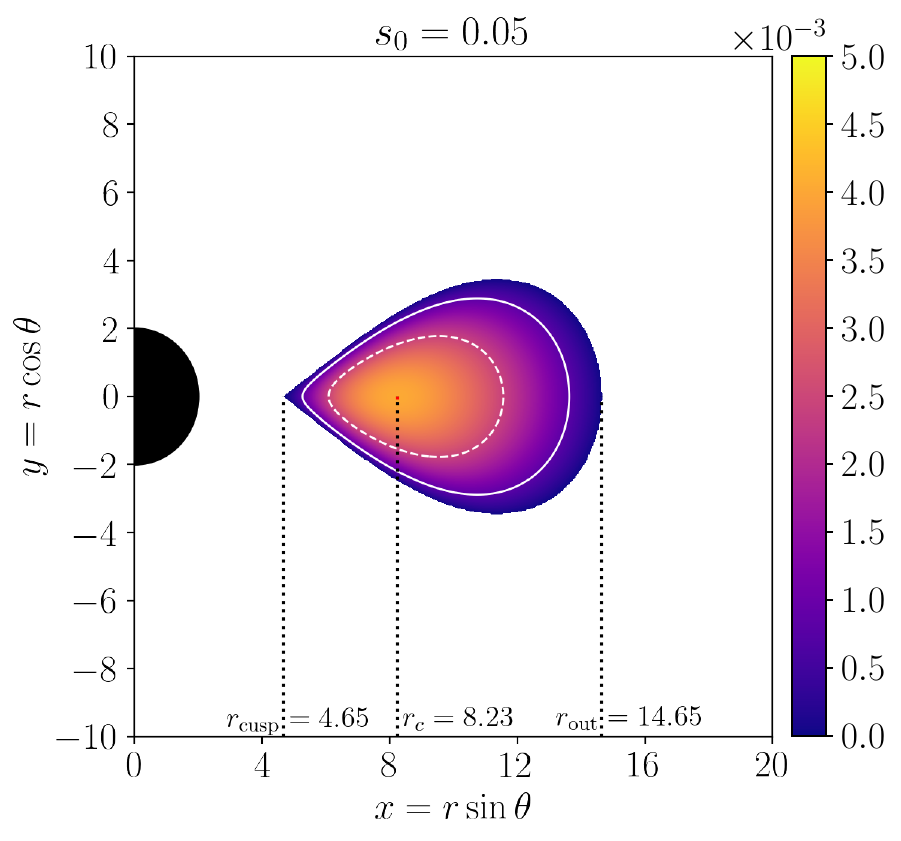}
  	\hspace{-0.1cm}
  		\\[3mm]
  		\hspace{-1.55cm}
  	\includegraphics[scale=0.52]{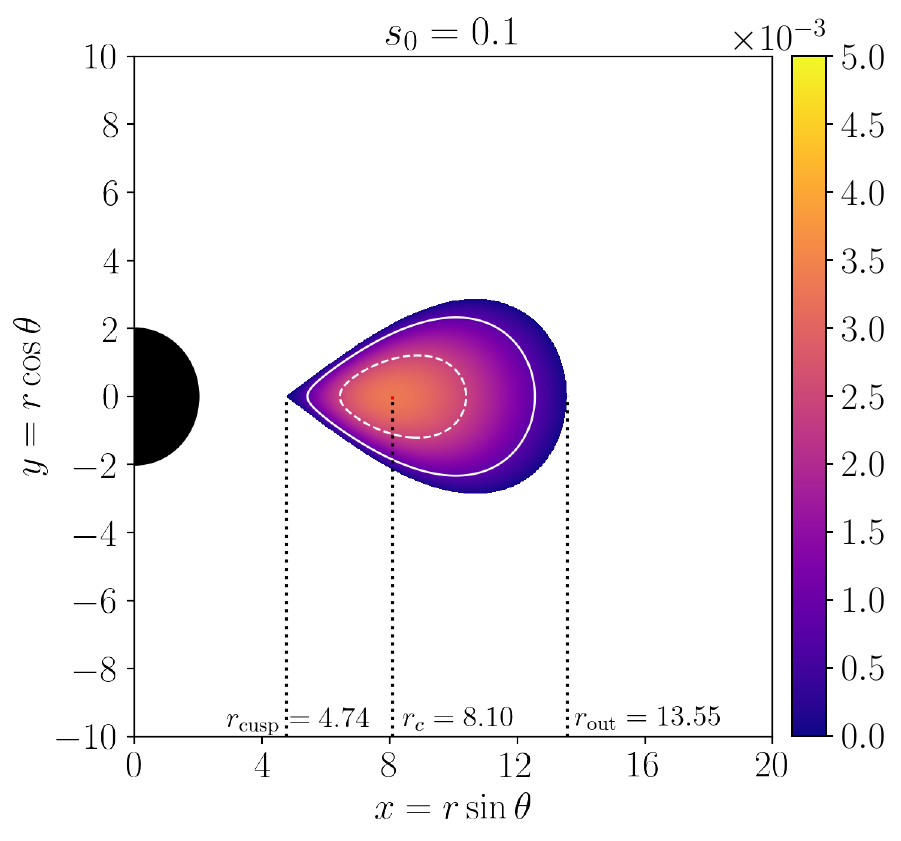}
  \hspace{-0.1cm}
  	\includegraphics[scale=0.52]{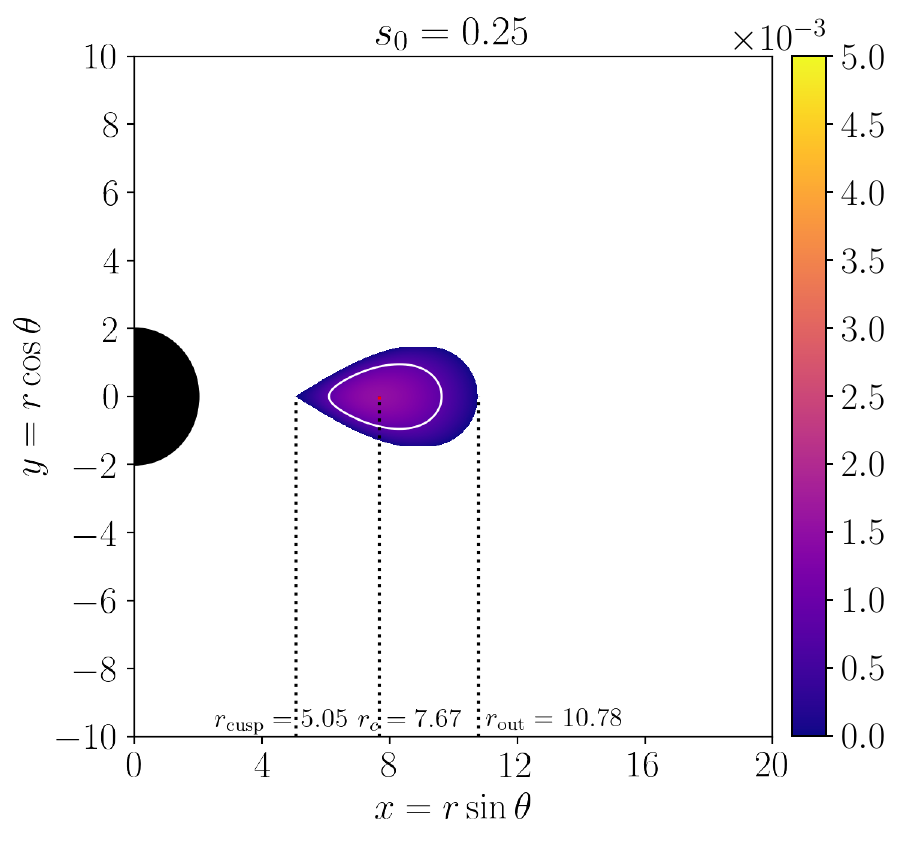}
  	\caption{Iso-density surfaces of a closed torus with $l_0=3.8$ and $s_0>0$. The small red dot depicts the center of the torus. To furnish a comparison with $s_0=0$ case, same constant density surfaces ( solid white and dotted white) are shown in all the panels. With increasing magnitude of $s_0$, the closed iso-density surfaces become smaller leading to an overall decrease in the size of the torus in comparison to a torus without a spin fluid. }  \label{fig2}
\end{figure*}

\begin{figure*}
\hspace{-1.45cm}
\captionsetup{justification=raggedright, singlelinecheck=on}
  \includegraphics[scale=0.52]{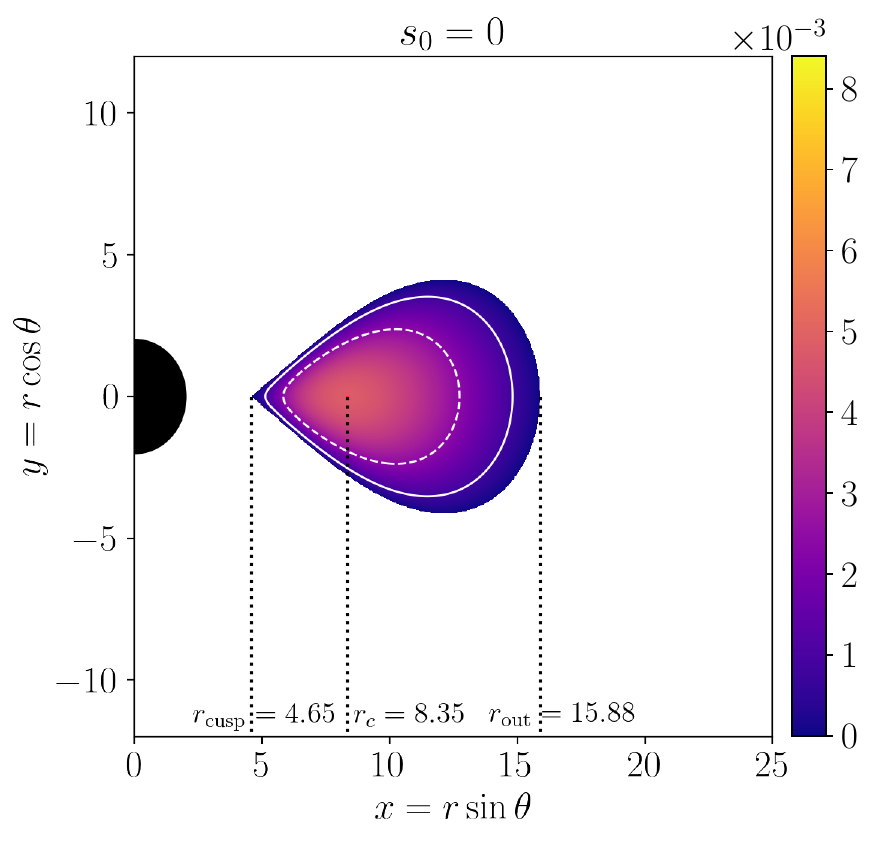}
  	\hspace{-0.1cm}
  	\includegraphics[scale=0.52]{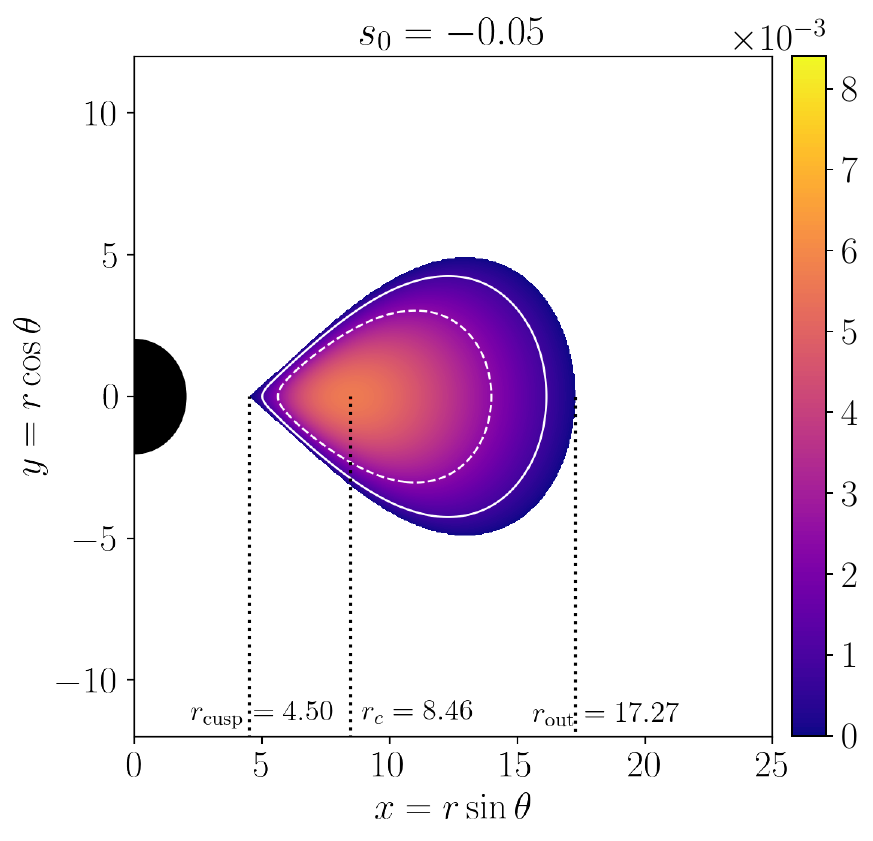}
  	\hspace{-0.1cm}
  		\\[3mm]
  		\hspace{-1.45cm}
  	\includegraphics[scale=0.52]{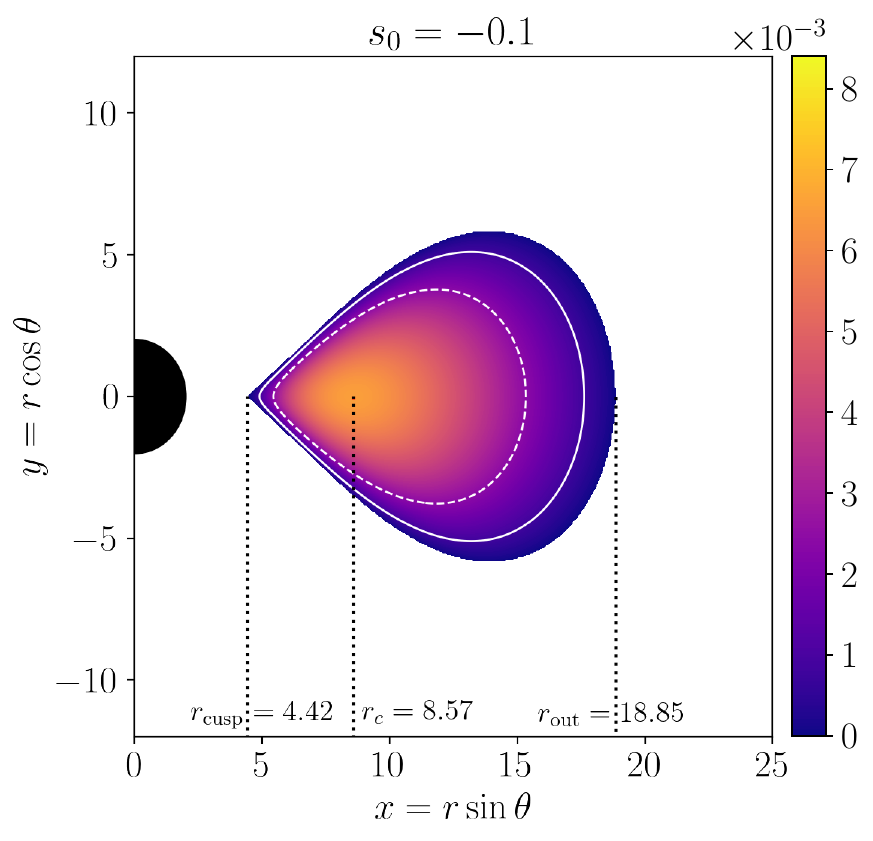}
  \hspace{-0.1cm}
  	\includegraphics[scale=0.52]{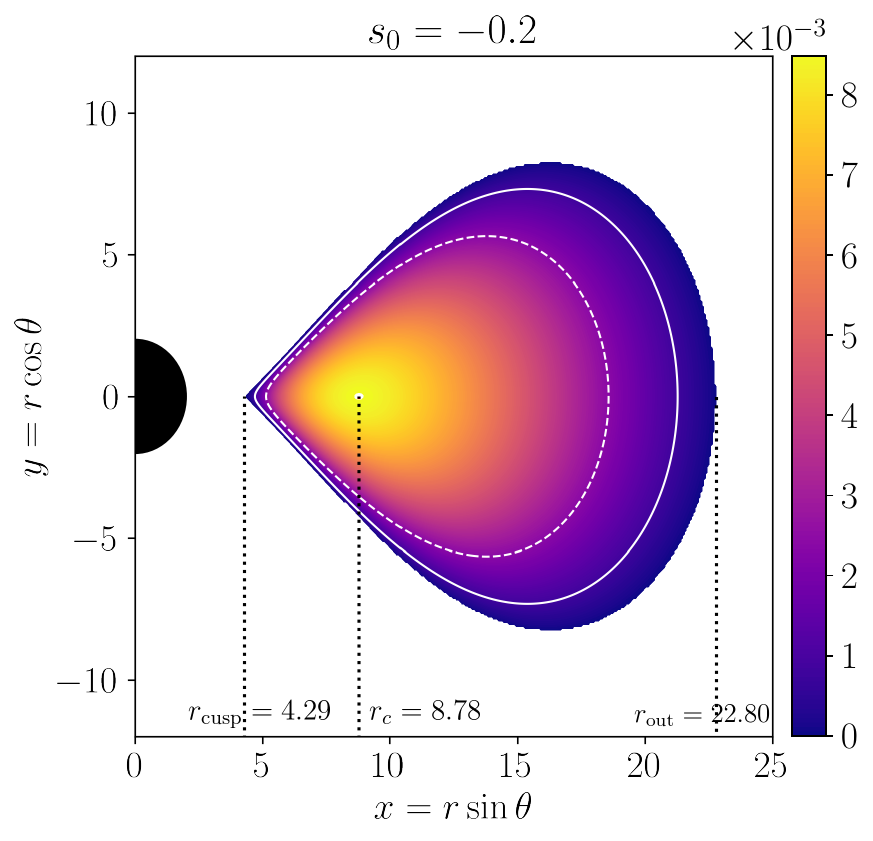}
  	\caption{Iso-density surfaces of the closed torus for $s_0<0$ and $l_0=3.8$.
   The small red dot depicts the center of the torus. To furnish a comparison with an ideal fluid torus, same constant density surfaces (solid white and dotted white lines) are shown in all the three panels which spread away leading to an increase in the size of the torus.} \label{fig3}
\end{figure*}

The implications of spin-curvature contributions on the morphology of a stationary geometrically thick spin fluid torus are illustrated by
the iso-energy density surfaces as shown in fig. \ref{fig2} for $s_0>0$ and fig. \ref{fig3} for $s_0<0$ 
In fig. \ref{fig2}, one finds that increasing $s_0$ leads to significant changes on the morphology of the torus.
First, the radial position of $r_{\mathrm{cusp}}$ shifts further away in comparison to $r_{\mathrm{cusp}}$ of the torus with $s=0$. At the same time, with further increase in $s_0$, $r_{\mathrm{out}}$ slowly moves closer to $r_{\mathrm{cusp}}$ resulting into decrease in the overall size of the torus which can be particularly noted for $s_0=0.25$. 
On the other hand, due to the spin and its coupling to curvature, the energy density at the center of the torus diminishes and the corresponding radial position slowly gets closer to the BH horizon. Such a feature is also observed in fig. \ref{fig2a} where the analysis is carried out at the equatorial plane. 
Second, the overall energy density of the torus gets systematically reduced with increasing magnitude of $s_0$. 
Third, the spin fluid contributes to redistribution of the iso-density surfaces which can be observed from fig. \ref{fig2} and  fig. \ref{fig3} respectively. For the purpose of illustration, it is observed that the reference contour lines $\epsilon_{\rm{max},0}/3$ (dashed white line) and $\epsilon_{\rm{max},0}/4$ (white line) ($\epsilon_{\rm{max},0}$ is the maximum energy density for $s_0=0$) shrink and become smaller with increasing $s_0$. 
\\
However, the scenario is exactly reverse in the $s_0<0$ region as shown in fig. \ref{fig3}. 
Due to the opposite alignment of spin (at the equatorial plane) the overall size of the torus gets enlarged resulting into a larger difference between $r_{\mathrm{cusp}}$ and $r_{\mathrm{out}}$, which can be particularly observed for $s_0=-0.2$ in fig. \ref{fig3}. 
In this case, the energy density at $r_c$ also increases while the location of $r_c$ shifts away from the horizon of the BH.
Notably, as $s_0$ is reduced further, the overall energy density of torus gets enhanced.
Similar to $s_0>0$ torus configurations, redistribution of constant density surfaces can be observed where the contour
lines expand increasing in the overall size of the torus.
\begin{figure*}
\captionsetup{justification=raggedright, singlelinecheck=on}
\hspace{-1.2cm}
  \includegraphics[scale=0.41]{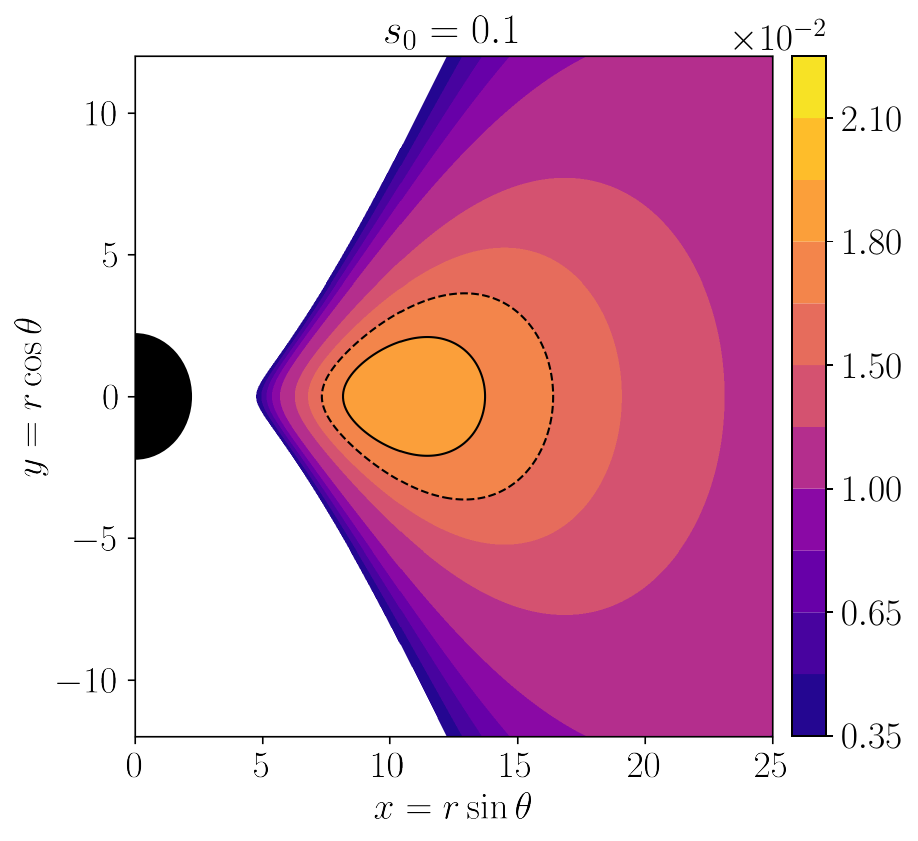}
  \hspace{-0.2cm}
  	\includegraphics[scale=0.41]{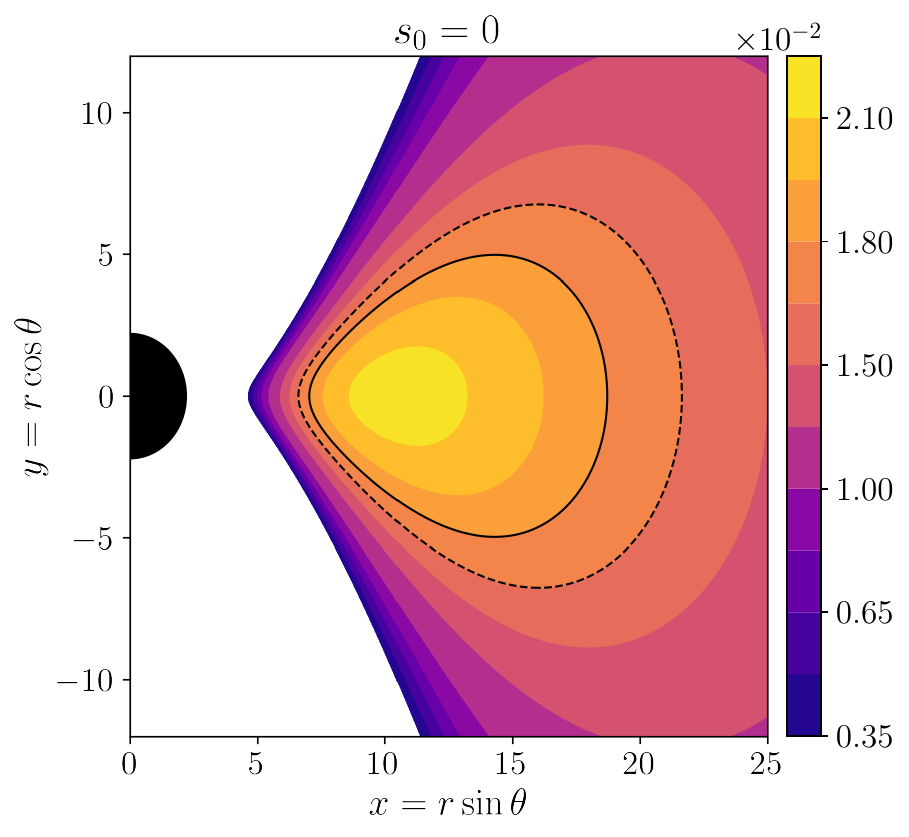}
  	\hspace{-0.2cm}
  		\includegraphics[scale=0.41]{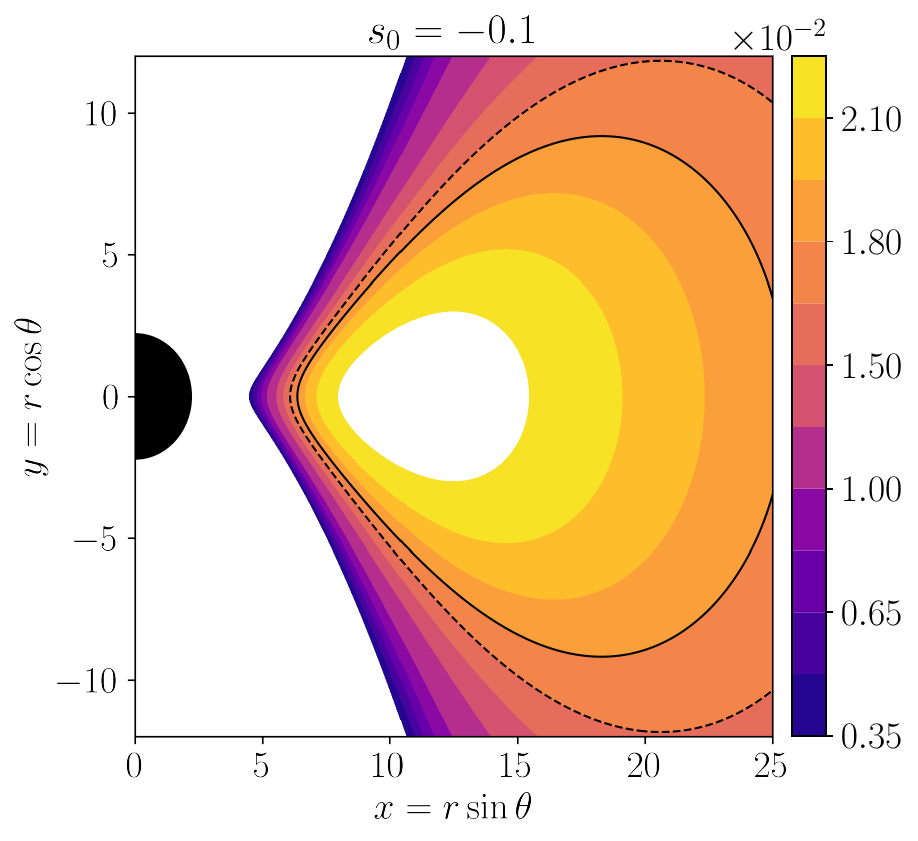}
\caption{Iso-density surfaces of an equilibrium torus corresponding to $s_0<0, s=0$, $s_>0$ with $l_0 = 4$. These tori are characterized by a cusp, center but are closed at infinity. The black solid and the black dotted lines illustrate redistribution of the same iso-density surfaces for three different values of $s_0$.  } \label{fig4}
\end{figure*}

By following a similar procedure as discussed before, the iso-density surfaces for $l_0=4$ are illustrated in fig. \ref{fig4} demonstrating a comparison between $s=0$ and $s_0 \neq 0$ tori.
These constant density surfaces possess a cusp and a center but are closed at infinity, a feature also found in a equilibrium solutions without a spin fluid in the Schwarzschild spacetime \cite{FoDai02}. Note that in presence of spin, the fluid lines move outwards for $s_0<0$ and inwards for $s_0>0$ in comparison to $s=0$. To highlight the spin-curvature effects and facilitate the comparison with the no-spin fluid torus configuration, we have plotted the same iso-density surfaces, namely  $\epsilon=0.0165$ (dotted black line) and $\epsilon=0.018$ (solid line) for all three cases. Similar to the case of $l_0=3.8$, the effects of spin can be observed from the movement of constant density surfaces which for $s_0>0$ come closer whereas move away for $s_0<0$ when compared with ideal fluid torus.

\begin{figure}[h]
\captionsetup{justification=raggedright, singlelinecheck=on}
    \centering
    \includegraphics[scale=0.47]{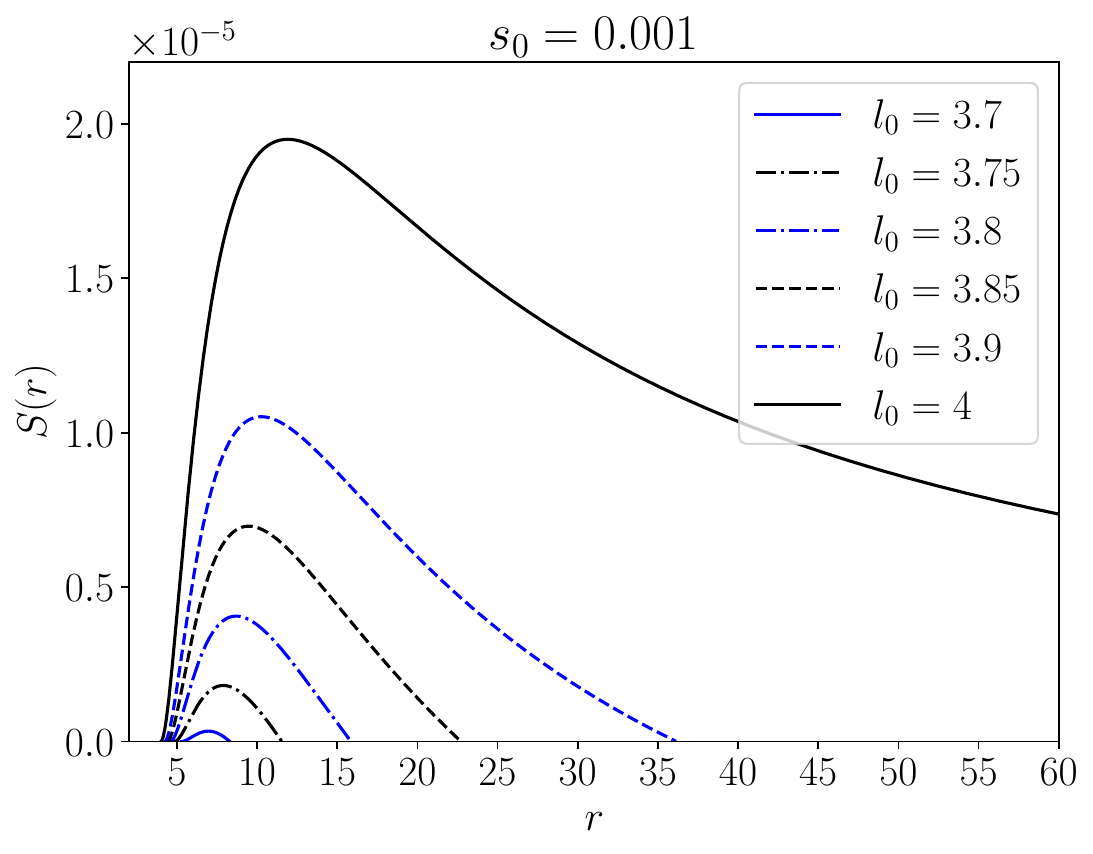}  
    \caption{Profiles of $S(r)$ at the equatorial plane for different values of $l_0$ and a fixed $s_0 = 0.001$.}
    \label{radSpin-1}
\end{figure}

The fig. \ref{radSpin-1} presents the profiles of spin length function at the equatorial plane for different values of $l_0$ and $s_0>0$. The figure shows that by increasing $l_0$, the peak value of $S(r)$ at the equatorial plane denoted by $S_{\mathrm{max}}$ increases further, the highest being for $l_0=4$, for which the torus is closed at infinity.  The corresponding radial position denoted by $r_{\mathrm{max}}$ is found to be steadily shifting away from the BH horizon. Furthermore, $r_\mathrm{out}$ also increases with $l_0$, implying an increase in size of the torus, which can be observed from the point of intersection of $S(r)$ with the x-axis.  In Table \ref{Tab1}, the solutions of $S_{\mathrm{max}}$ and $r_{\mathrm{max}}$ for a representative value of $s_0 =0.001$ are presented , from which, it is readily observed $r_{\mathrm{max}}$ never coincides with $r_c$.
On the other hand, for $s_0<0$ the spin length function takes negative values due to negative sign of $s_0$. Once again, similar to $s_0>0$ case, $r_{\mathrm{max}}$ does not coincide with $r_c$ while ($S_{\mathrm{max}}$ and $r_{\mathrm{out}}$ decreases with $l_0$).
In Table \ref{Tab2}, representative values of $l_0$ (which generates solutions of equilibrium tori) are chosen to illustrate the behaviour of $S_{\mathrm{max}}$ and $r_{\mathrm{max}}$ due to changing values of $s_0$. For a fixed $l_0$, $S_{\mathrm{max}}$ increases but $r_{\mathrm{max}}$ decreases in $s_0>0$ region. This picture is opposite for all $s_0<0$ cases, where one finds that $S_{\mathrm{max}}$ takes a dip and $r_{\mathrm{max}}$ increases in the similar way as $r_c$.
Table \ref{Tab3} shows a series of equilibrium models of closed spin fluid torus for both positive and negative values of $s_0$.
\begin{table*}[h!]
   \begin{tabular*}{0.32\linewidth}{ |p{0.95cm}|p{1.2cm}|p{1.8cm}|p{1.2cm}|}
    \hline 
         \multicolumn{4}{|c|}{$ s_0 = 0.001$} \\
         \hline 
          $l_0$ & $r_{\rm c}$, & $S_{\mathrm{max}}$ &  $r_{\mathrm{max}}$ \\
          \hline
          3.7 & 6.91  & $3.73 \times 10^{-7}$& 6.96 \\
          \hline
          3.75 & 7.71  & $1.81 \times 10^{-6}$& 7.87 \\
          \hline
          3.8 & 8.35  & $4.06 \times 10^{-6}$& 8.69 \\
          \hline
          3.85 & 8.91  & $6.97 \times 10^{-6}$& 6.97 \\
          \hline
          3.9 & 9.45  & $1.05 \times 10^{-5}$& 10.25 \\
          \hline
          4 & 10.47  & $1.95 \times 10^{-5}$& 11.99 \\
          \hline
          \end{tabular*}
 \caption{ Variation of $S_{\mathrm{max}}$ and $r_{\mathrm{max}}$ for different values of $l_0$ and fixed $s_0$.}
  \label{Tab1}
\end{table*}

\begin{table*}[ht]
\centering
   \begin{tabular*}{0.635\linewidth}
   { |p{1cm}|p{1.1cm}|p{1.8cm}|p{1.1cm}||p{1cm}|p{1.1cm}|p{2cm}|p{1.1cm}|}
    \hline 
         \multicolumn{8}{|c|}{$l_0 =3.8$, $\kappa =1$} \\
         \hline 
          $s_0$ & $r_{\rm c}$, & $S_{\mathrm{max}}$ &  $r_{\mathrm{max}}$ & $s_0$ & $r_{\rm c}$ & $S_{\mathrm{max}}$ &  $r_{\mathrm{max}}$ \\
          \hline
          0.001 & 8.35  & $4.06 \times 10^{-6}$& 8.70 & -0.001 & 8.35  & $-4.09 \times 10^{-6}$& 8.71 \\
          \hline
          0.01 & 8.32  & $3.93 \times 10^{-5}$& 8.67
          & -0.01 & 8.37  & $-4.21 \times 10^{-5}$& 8.73\\
          \hline
          0.1 & 8.10  & $2.78 \times 10^{-4}$& 8.36 
          & -0.1 & 8.57  & $-5.55 \times 10^{-4}$& 9.03\\
          \hline
          0.25 & 7.67  & $3.11 \times 10^{-4}$& 7.81 
          & -0.25 & 8.88  & $-2.06 \times 10^{-3}$& 9.51\\
          \hline
          0.35 & 7.31  & $1.17 \times 10^{-4}$& 7.38 & -0.35 & 9.07  & $-3.62 \times 10^{-3}$& 9.82 \\
          \hline
          0.4 & 7.07  & $1\times 10^{-4}$& 7.12
          & -0.4 & 9.16 & $-4.58 \times 10^{-3}$& 9.98 \\
          \hline
          \end{tabular*}
 \caption{ The stationary solutions of an ideal Weyssenhoff fluid torus showing the behaviour of $S_{\mathrm{max}}$ and $r_{\mathrm{max}}$ with $s_0$ for both $s_0>0$ and $s_0<0$. Note that $r_{\mathrm{max}}$ never coincides with center of the torus $r_{c}$.}
  \label{Tab2}
\end{table*}

\begin{table*}[ht!]
\centering
   \begin{tabular*}{0.99\linewidth}{ |p{1cm}|p{1.2cm}|p{1.2cm}|p{1.6cm}|p{2cm}|p{1cm}|p{1cm}|p{1.2cm}|p{1.2cm}|p{1.6cm}|p{2cm}|p{1cm}|}
    \hline 
         \multicolumn{12}{|c|}{$S(r,\theta)=s_0  k(r,\theta)(1-\Omega l_0)\epsilon(r,\theta),\quad l_0 = 3.8, \quad \kappa =1$} \\
         \hline 
          $s_0$ & $r_{\rm cusp}$ & $r_{c}$ & $\epsilon_{\rm max}$ & $p_{\rm max}$ & $r_{\rm out}$ & $s_0$ & $r_{\rm cusp}$ & $r_{c}$ & $\epsilon_{\rm max}$ & $p_{\rm max}$ & $r_{\rm out}$  \\
          \hline
          0 & 4.57 & 8.35 & 0.0048 & 2.31 $\times$ $10^{-5}$ & 15.89 & 0 & 4.57 & 8.35 & 0.0048 & 2.31 $\times$ $10^{-5}$ & 15.89 \\
         \hline
          0.01 & 4.59 & 8.32 & 0.0046 & $2.16 \times 10^{-5}\,$ & 15.63 & $-0.01$ & 4.56 & 8.37 & 0.0049 & 2.47 $\times$ $10^{-5}$ & 16.15 \\
         \hline
          0.05 & 4.65 & 8.23 & 0.0040 & $1.62 \times 10^{-5}\,$ & 15.45 & $-0.05$ & 4.50 & 8.46 & 0.0056 & 3.18 $\times$ $10^{-5}$ & 17.27 \\
         \hline
          0.15 & 4.83 & 7.97 & 0.0026 & $7.04 \times 10^{-6}\,$ & 12.54 & $-0.15$ & 4.36 & 8.68 & 0.0074 & 5.55 $\times$ $10^{-5}$ & 20.67\\
         \hline
          0.25 & 5.05 & 7.67 & 0.0015 & $2.29 \times 10^{-6}\,$ & 10.75 & $-0.25$ & 4.23 & 8.88 & 0.0095 & 9.12 $\times$ $10^{-5}$ & 25.33 \\
         \hline
          0.35 & 5.33 & 7.31 & 0.00063 & $4.04 \times 10^{-7}\,$ & 9.14 & $-0.35$ & 4.12 & 9.07 & 0.0118 & 1.40 $\times$ $10^{-4}$ & 32.21 \\
         \hline
          0.45 & 5.78 & 6.76 & $7.52 \times 10^{-5}$& $5.65 \times 10^{-9}\,$ & 7.41 & $-0.45$ & 4.02 &  9.25 & 0.0144 & 2.07 $\times$ $10^{-4}$ & 43.50\\
         \hline 
    \end{tabular*}
 \caption{Solutions of Stationary equilibrium model closed torus supported by ideal Weyssenhoff spin fluid for $s_0\neq 0$. The solutions are compared with purely ideal fluid torus charecterized by  $s_0 =0$.  }
  \label{Tab3}
\end{table*}

Since the Schwarzschild BH spacetime and the Weyssenhoff ideal spin fluid share the same symmetries, in congruence to the orientation of spin vector at the equatorial plane,  the non-zero contributions of the spin-curvature coupling term only come from $R_{trtr} S^{tr}u^{t}$ and $R_{r\phi \phi r} S^{r \phi}u^{\phi}$.
Together with \eqref{compspin1}, \eqref{compspin2} and \eqref{energy1}) we obtain, 
\begin{eqnarray}
    R_{trtr} S^{tr} u^{t}
   =   \frac{2 l_0 s_0}{r^4 \sin \theta} k(r, \theta)\epsilon(r,\theta) \nonumber \\
    = \frac{2 l_0 s_0 k}{r^4 } \left[\frac{2^{1/4}\kappa C_1(r)(r^3 \sin^2 \theta-l_0^2(r-2))^{1/4}-\sqrt{\sin \theta}}{\kappa \sin ^{3/2}\theta}\right] \label{compspin3} \nonumber \\
    \end{eqnarray}
and
\begin{eqnarray}
    R_{r\phi \phi r} S^{r\phi} u^{\phi} =  \frac{ l_0 s_0 k(r, \theta)}{r^4 \sin \theta} \epsilon(r,\theta) \nonumber \\
    = \frac{l_0 s_0 k }{r^4}\left[\frac{2^{1/4}\kappa C_1(r)(r^3 \sin^2 \theta-l_0^2(r-2))^{1/4}-\sqrt{\sin \theta}}{\kappa \sin ^{3/2}\theta}\right] \label{compspin4}\nonumber \\
\end{eqnarray}
where $k(r, \theta)$ is given by \eqref{soln-k} and the constant of integration $C_1(r)$ is determined at the equatorial plane. 
It is observed that the non-zero components of the spin-curvature coupling term given by \eqref{compspin3} and \eqref{compspin4} are symmetric in both upper and lower hemispheres under our present considerations which therefore indicate a symmetric contribution of spin in both the hemispheres of the torus. 
Both fig. \ref{fig2} and fig. \ref{fig3} confirm this fact about the stationary equilibrium solutions of the torus obtained with $s_0\neq 0$ in presence of spin.

The stationary solutions of equilibrium torus with ideal Weyssenhoff spin fluid demonstrate relocation of radial positions of the cusp and center with changing magnitudes of $s_0$. The two radial positions correspond to crossing of the actual specific angular momentum and the Keplerian specific angular momentum which are determined from the condition $\displaystyle\frac{\partial_{\mu}p}{\epsilon+p}=0=a_{\mu}$.  In presence of spin, it is expected the Keplerian specific angular momentum will be modified as well \cite{1979MNRAS.189..621A} and is determined from the corresponding crossing condition.   By putting  $\partial_{r} \epsilon =0$ in \eqref{eq:15} together with \eqref{energy1}, the Keplerian specific angular momentum for the spin fluid is then obtained by solving following equation \footnote{We thank M.Abramowciz for bringing the reference \cite{1979MNRAS.189..621A} to our notice and suggesting the computation.},
\begin{eqnarray}
    [r^3-l_0^2(r-2)]^{1/4}+\frac{2^{3/4}l_0 s_0(r-3)[r^3-l_0^2(r-2)^2]}{C_2(r)r^4(r-2)(r^3-l_0^2(r-2))}=0 \label{kep}\nonumber \\
\end{eqnarray}
One can easily observe that in absence of spin, the above relation gives rise to $l_0=l_k$. For a given $s_0$, Keplerian specific angular momentum is determined by solving \eqref{kep} for $l_0$ after substituting $C_2(r)$ from \eqref{C2-eqn1} . 





\section{Estimation of the spin parameter}
The spin function $S(r, \theta)$ plays a crucial role in determining the effects of spin on the equilibrium solutions of a torus. Therefore in order to pinpoint the spin effects for astrophysical objects, it is important to estimate the relevant magnitude of spin. 

In geometrised units, length, time and mass have same dimensions. In these units, the dimension of the spin tensor is  given by $S^{\alpha \beta}$ =$[L]$ using (\ref{mom-eqn}) whose non-zero components of spin tensor are given by Eq. \ref{compspin1} and  Eq. \ref{compspin2}. The dimension of the spin length function is given by $[S]=[L]$. 
Note that $\epsilon$, $k$ are dimensionless in geometrized units, from the relation, $S=s_0 \epsilon k (1-\Omega l_0)$, one readily finds the dimension of $s_0$ as  $[s_0]=[L]$.

In order to have a rough estimation of $s_0$, let us assume that the torus is composed of electrons only. The specific spin (spin per unit mass) $\bar{S}$ of an electron is $(\hbar/2)/m_e = 1.93 \times 10^{-11} $ cm, where $m_e$ is the mass of an electron \cite{Semerak:1999qc}. In terms of units of mass of a central compact object with $M=3 M_\odot =4.43 \times 10^{5}$ cm, the specific spin is given by $\bar{S}=0.44 \times 10^{-16}m_{e}M$. 
On the other hand, we can express the specific spin as $\bar{S}=s_0k(r,\theta)$. On comparing, $s_0=0.44 \times 10^{-16}m_{e}M$, where the value of $k(r,\theta)=1$ is taken at the equatorial plane.
The electron density within an accretion disk $n=10^{18}-10^{23}$ cm$^3$. With this, $s_0 \approx 10^{2}- 10^{5}$ \cite{2020AAS...23534604G}. On the other hand, if the central compact object is a supermassive BH with mass $M=10^{8} M_\odot$, $s_0=10^{-6}-0.1\, m_e M$cm$^{-3}$. 


\section{Summary and Outlook}
The study of non-self gravitating ideal fluid moving in circular motion under hydrostatic equilibrium in a BH spacetime  has remained a topic of active research over a long time, thanks to the simple theoretical framework that can be used for developing analytical stationary equilibrium solutions, which serve as initial data for developing numerical simulations of realistic accretion flows. 
In this work, we have considered a different matter model where the the fluid constituents carry an additional intrinsic angular momentum, proportional to their volume. 
In this case, the spin angular momentum density of the fluid is described by the rank-two spin tensor which is related to the macroscopic spin vector (see, \eqref{sv} ) and the matter model of the torus is described by the neutral ideal Weyssenhoff spin fluid.  Subjected to Frenkel SSC, the corresponding momentum balance equation is characterized by its spin and their coupling to curvature of the Schwarzschild BH, all of which have substantial effects on stationary solutions of the equilibrium torus. 

By further allowing the spin fluid to undergo circular motion in a stationary and axisymmetric background and taking the orientation of the macroscopic spin vector only perpendicular to the equatorial plane, the present work addresses following issues. 
First, we have determined the integrability conditions of the general relativistic momentum conservation equation using the Frenkel SSC. 
These new set of integrability conditions are found to be straight forward extension of Von-Zeipel conditions with additional conditions 
that emerge solely due to the spin, spacetime curvature of a compact object and their couplings. 
Although integrability conditions of spinning \textit{particles} have been earlier addressed (see, for example, \cite{Lukes-Gerakopoulos:2016udm}), to the best of our knowledge, the integrability conditions of a spin \textit{fluid} moving in stationary and axisymmetric spacetime, are, for the first time, reported in the present work.
Second, the existence of integrability conditions has allowed us to determine the spin length function (related to the spin tensor) and hence the complete structure of the spin tensor. Third, using the spin tensor, the general relativistic momentum conservation equation is solved for constructing stationary models of equilibrium tori in Schwarzschild geometry by assuming constant specific angular momentum distributions and a polytropic equation of state.

Set in this simple scenario, this work presents several novel features as described.
On comparing with purely ideal fluid torus, our study reveals that the morphology of the torus is significantly modified in presence of the coupling term of the spin and the curvature tensor of the central BH. 
The most pronounced signature of spin is found to be for the case of closed torus configurations.  Therefore to furnish a quantitative study for determining the effects of spin on stationary solutions of the torus, the spin length 
is assigned with a constant free parameter $s_0$, that is allowed to take both positive and negative values characterising alignment/anti-alignment of the spin vector perpendicular to the equatorial plane. Then, $s_0$ is varied systematically from $s=0$ to both higher positive and negative values to analyse the consequences of spin.\\
Our analysis reveals that, increasing $s_0$ from $s_0=0$ leads to decrease in the overall energy density and pressure of the torus.  The size of the torus also decreases which can be noticed from the shift of $r_\mathrm{cusp}$ away from the BH horizon and $r_{\mathrm{out}}$ towards the horizon. On the other hand, the separation distance between $r_\mathrm{cusp}$ and $r_{\mathrm{out}}$ 
increase with systematic increase in $s_0$ for anti-alignment of the spin, leading to enhancement of both the energy density and the pressure while also expanding its size.  
In addition, the center of the torus gets shifted, either away or towards the horizon depending on $s_0$ in comparison to an ideal spin fluid torus. 
Eventually, the maximum energy density at the center of the torus decreases (for tori with $s_0>0$) or increases (for tori with $s_0>0$).
It is further noted that the existence of additional intrinsic spin angular momentum leads to redistribution of iso-density contours in the torus configurations. 
On studying independently, the behaviour of spin function at the equatorial plane  with respect to changes in $l_0$ and $s_0$ it is found the radial location corresponding to its maximum value is shifted for all $s\neq 0$ and hence never coincides with the torus center. 
Finally, the value of $s_0$ is estimated by considering the torus to be composed of spin 1/2 particles.

The present work is the first step towards studying the stationary solutions of geometrically thick tori in presence of ideal Weyssenhoff spin fluid. We adopted a several assumptions and presented a simple model, nevertheless, it is found that presence of additional intrinsic angular momentum of fluid elements in the form of spin produces a number of new features on the morphology of a torus. Given the integrability conditions in hand, it will be interesting to investigate the possible changes of the morphology of the torus situated in a rotating background, for example, in the astrophysically relevant Kerr BH spacetime. The other realistic case worth exploring is the case of non-constant specific angular momentum distributions of the torus. Particularly, for realistic accretion flow, when a viscous effects are no longer small to be neglected, it would be pertinent to consider shear viscous effects within the spin fluid, for example, using a causal prescription \cite{Lahiri:2019mwc}. We leave all these issues for future investigations.

\section*{Acknowledgement}
We are thankful to Marek Abramowicz and Vladimir Karas for carefully going through the manuscript and providing us with insightful comments. The work of SL is supported by the Deutsche Forschungsgemeinschaft (DFG) with grant number 4040115. 
\bibliography{accretion3}

 \newcommand{\noop}[1]{}
\begin{thebibliography}{62}%
\makeatletter
\providecommand \@ifxundefined [1]{%
 \@ifx{#1\undefined}
}%
\providecommand \@ifnum [1]{%
 \ifnum #1\expandafter \@firstoftwo
 \else \expandafter \@secondoftwo
 \fi
}%
\providecommand \@ifx [1]{%
 \ifx #1\expandafter \@firstoftwo
 \else \expandafter \@secondoftwo
 \fi
}%
\providecommand \natexlab [1]{#1}%
\providecommand \enquote  [1]{``#1''}%
\providecommand \bibnamefont  [1]{#1}%
\providecommand \bibfnamefont [1]{#1}%
\providecommand \citenamefont [1]{#1}%
\providecommand \href@noop [0]{\@secondoftwo}%
\providecommand \href [0]{\begingroup \@sanitize@url \@href}%
\providecommand \@href[1]{\@@startlink{#1}\@@href}%
\providecommand \@@href[1]{\endgroup#1\@@endlink}%
\providecommand \@sanitize@url [0]{\catcode `\\12\catcode `\$12\catcode
  `\&12\catcode `\#12\catcode `\^12\catcode `\_12\catcode `\%12\relax}%
\providecommand \@@startlink[1]{}%
\providecommand \@@endlink[0]{}%
\providecommand \url  [0]{\begingroup\@sanitize@url \@url }%
\providecommand \@url [1]{\endgroup\@href {#1}{\urlprefix }}%
\providecommand \urlprefix  [0]{URL }%
\providecommand \Eprint [0]{\href }%
\providecommand \doibase [0]{http://dx.doi.org/}%
\providecommand \selectlanguage [0]{\@gobble}%
\providecommand \bibinfo  [0]{\@secondoftwo}%
\providecommand \bibfield  [0]{\@secondoftwo}%
\providecommand \translation [1]{[#1]}%
\providecommand \BibitemOpen [0]{}%
\providecommand \bibitemStop [0]{}%
\providecommand \bibitemNoStop [0]{.\EOS\space}%
\providecommand \EOS [0]{\spacefactor3000\relax}%
\providecommand \BibitemShut  [1]{\csname bibitem#1\endcsname}%
\let\auto@bib@innerbib\@empty
\bibitem [{\citenamefont {Oppenheimer}\ and\ \citenamefont
  {Snyder}(1939)}]{PhysRev.56.455}%
  \BibitemOpen
  \bibfield  {author} {\bibinfo {author} {\bibfnamefont {J.~R.}\ \bibnamefont
  {Oppenheimer}}\ and\ \bibinfo {author} {\bibfnamefont {H.}~\bibnamefont
  {Snyder}},\ }\bibfield  {title} {\enquote {\bibinfo {title} {On continued
  gravitational contraction},}\ }\href {\doibase 10.1103/PhysRev.56.455}
  {\bibfield  {journal} {\bibinfo  {journal} {Phys. Rev.}\ }\textbf {\bibinfo
  {volume} {56}},\ \bibinfo {pages} {455--459} (\bibinfo {year}
  {1939})}\BibitemShut {NoStop}%
\bibitem [{\citenamefont {Penrose}(1965)}]{PhysRevLett.14.57}%
  \BibitemOpen
  \bibfield  {author} {\bibinfo {author} {\bibfnamefont {Roger}\ \bibnamefont
  {Penrose}},\ }\bibfield  {title} {\enquote {\bibinfo {title} {Gravitational
  collapse and space-time singularities},}\ }\href {\doibase
  10.1103/PhysRevLett.14.57} {\bibfield  {journal} {\bibinfo  {journal} {Phys.
  Rev. Lett.}\ }\textbf {\bibinfo {volume} {14}},\ \bibinfo {pages} {57--59}
  (\bibinfo {year} {1965})}\BibitemShut {NoStop}%
\bibitem [{\citenamefont {Kormendy}\ and\ \citenamefont
  {Ho}(2013)}]{Kormendy:2013dxa}%
  \BibitemOpen
  \bibfield  {author} {\bibinfo {author} {\bibfnamefont {John}\ \bibnamefont
  {Kormendy}}\ and\ \bibinfo {author} {\bibfnamefont {Luis~C.}\ \bibnamefont
  {Ho}},\ }\bibfield  {title} {\enquote {\bibinfo {title} {{Coevolution (Or
  Not) of Supermassive Black Holes and Host Galaxies}},}\ }\href {\doibase
  10.1146/annurev-astro-082708-101811} {\bibfield  {journal} {\bibinfo
  {journal} {Ann. Rev. Astron. Astrophys.}\ }\textbf {\bibinfo {volume} {51}},\
  \bibinfo {pages} {511--653} (\bibinfo {year} {2013})},\ \Eprint
  {http://arxiv.org/abs/1304.7762} {arXiv:1304.7762 [astro-ph.CO]} \BibitemShut
  {NoStop}%
\bibitem [{\citenamefont {Falcke}\ \emph {et~al.}(2000)\citenamefont {Falcke},
  \citenamefont {Melia},\ and\ \citenamefont {Agol}}]{Falcke:1999pj}%
  \BibitemOpen
  \bibfield  {author} {\bibinfo {author} {\bibfnamefont {Heino}\ \bibnamefont
  {Falcke}}, \bibinfo {author} {\bibfnamefont {Fulvio}\ \bibnamefont {Melia}},
  \ and\ \bibinfo {author} {\bibfnamefont {Eric}\ \bibnamefont {Agol}},\
  }\bibfield  {title} {\enquote {\bibinfo {title} {{Viewing the shadow of the
  black hole at the galactic center}},}\ }\href {\doibase 10.1086/312423}
  {\bibfield  {journal} {\bibinfo  {journal} {Astrophys. J. Lett.}\ }\textbf
  {\bibinfo {volume} {528}},\ \bibinfo {pages} {L13} (\bibinfo {year}
  {2000})},\ \Eprint {http://arxiv.org/abs/astro-ph/9912263}
  {arXiv:astro-ph/9912263} \BibitemShut {NoStop}%
\bibitem [{\citenamefont {Narayan}\ and\ \citenamefont
  {McClintock}(2013)}]{Narayan:2013gca}%
  \BibitemOpen
  \bibfield  {author} {\bibinfo {author} {\bibfnamefont {Ramesh}\ \bibnamefont
  {Narayan}}\ and\ \bibinfo {author} {\bibfnamefont {Jeffrey~E.}\ \bibnamefont
  {McClintock}},\ }\bibfield  {title} {\enquote {\bibinfo {title}
  {{Observational Evidence for Black Holes}},}\ }\href@noop {} {\  (\bibinfo
  {year} {2013})},\ \Eprint {http://arxiv.org/abs/1312.6698} {arXiv:1312.6698
  [astro-ph.HE]} \BibitemShut {NoStop}%
\bibitem [{\citenamefont {Akiyama}\ \emph {et~al.}(2019)\citenamefont {Akiyama}
  \emph {et~al.}}]{EventHorizonTelescope:2019dse}%
  \BibitemOpen
  \bibfield  {author} {\bibinfo {author} {\bibfnamefont {Kazunori}\
  \bibnamefont {Akiyama}} \emph {et~al.} (\bibinfo {collaboration} {Event
  Horizon Telescope}),\ }\bibfield  {title} {\enquote {\bibinfo {title} {{First
  M87 Event Horizon Telescope Results. I. The Shadow of the Supermassive Black
  Hole}},}\ }\href {\doibase 10.3847/2041-8213/ab0ec7} {\bibfield  {journal}
  {\bibinfo  {journal} {Astrophys. J. Lett.}\ }\textbf {\bibinfo {volume}
  {875}},\ \bibinfo {pages} {L1} (\bibinfo {year} {2019})},\ \Eprint
  {http://arxiv.org/abs/1906.11238} {arXiv:1906.11238 [astro-ph.GA]}
  \BibitemShut {NoStop}%
\bibitem [{\citenamefont {Akiyama}\ \emph {et~al.}(2022)\citenamefont {Akiyama}
  \emph {et~al.}}]{EventHorizonTelescope:2022xnr}%
  \BibitemOpen
  \bibfield  {author} {\bibinfo {author} {\bibfnamefont {Kazunori}\
  \bibnamefont {Akiyama}} \emph {et~al.} (\bibinfo {collaboration} {Event
  Horizon Telescope}),\ }\bibfield  {title} {\enquote {\bibinfo {title} {{First
  Sagittarius A* Event Horizon Telescope Results. I. The Shadow of the
  Supermassive Black Hole in the Center of the Milky Way}},}\ }\href {\doibase
  10.3847/2041-8213/ac6674} {\bibfield  {journal} {\bibinfo  {journal}
  {Astrophys. J. Lett.}\ }\textbf {\bibinfo {volume} {930}},\ \bibinfo {pages}
  {L12} (\bibinfo {year} {2022})}\BibitemShut {NoStop}%
\bibitem [{\citenamefont {Abbott}\ \emph {et~al.}(2016)\citenamefont {Abbott}
  \emph {et~al.}}]{LIGOScientific:2016aoc}%
  \BibitemOpen
  \bibfield  {author} {\bibinfo {author} {\bibfnamefont {B.~P.}\ \bibnamefont
  {Abbott}} \emph {et~al.} (\bibinfo {collaboration} {LIGO Scientific,
  Virgo}),\ }\bibfield  {title} {\enquote {\bibinfo {title} {{Observation of
  Gravitational Waves from a Binary Black Hole Merger}},}\ }\href {\doibase
  10.1103/PhysRevLett.116.061102} {\bibfield  {journal} {\bibinfo  {journal}
  {Phys. Rev. Lett.}\ }\textbf {\bibinfo {volume} {116}},\ \bibinfo {pages}
  {061102} (\bibinfo {year} {2016})},\ \Eprint
  {http://arxiv.org/abs/1602.03837} {arXiv:1602.03837 [gr-qc]} \BibitemShut
  {NoStop}%
\bibitem [{\citenamefont {{Webster}}\ and\ \citenamefont
  {{Murdin}}(1972)}]{1972Natur.235...37W}%
  \BibitemOpen
  \bibfield  {author} {\bibinfo {author} {\bibfnamefont {B.~Louise}\
  \bibnamefont {{Webster}}}\ and\ \bibinfo {author} {\bibfnamefont {Paul}\
  \bibnamefont {{Murdin}}},\ }\bibfield  {title} {\enquote {\bibinfo {title}
  {{Cygnus X-1-a Spectroscopic Binary with a Heavy Companion ?}}}\ }\href
  {\doibase 10.1038/235037a0} {\bibfield  {journal} {\bibinfo  {journal}
  {\nat}\ }\textbf {\bibinfo {volume} {235}},\ \bibinfo {pages} {37--38}
  (\bibinfo {year} {1972})}\BibitemShut {NoStop}%
\bibitem [{\citenamefont {{McClintock}}\ and\ \citenamefont
  {{Remillard}}(1986)}]{1986ApJ...308..110M}%
  \BibitemOpen
  \bibfield  {author} {\bibinfo {author} {\bibfnamefont {J.~E.}\ \bibnamefont
  {{McClintock}}}\ and\ \bibinfo {author} {\bibfnamefont {R.~A.}\ \bibnamefont
  {{Remillard}}},\ }\bibfield  {title} {\enquote {\bibinfo {title} {{The Black
  Hole Binary A0620-00}},}\ }\href {\doibase 10.1086/164482} {\bibfield
  {journal} {\bibinfo  {journal} {\apj}\ }\textbf {\bibinfo {volume} {308}},\
  \bibinfo {pages} {110} (\bibinfo {year} {1986})}\BibitemShut {NoStop}%
\bibitem [{\citenamefont {{Frank}}\ \emph {et~al.}(2002)\citenamefont
  {{Frank}}, \citenamefont {{King}},\ and\ \citenamefont
  {{Raine}}}]{2002apa..book.....F}%
  \BibitemOpen
  \bibfield  {author} {\bibinfo {author} {\bibfnamefont {Juhan}\ \bibnamefont
  {{Frank}}}, \bibinfo {author} {\bibfnamefont {Andrew}\ \bibnamefont
  {{King}}}, \ and\ \bibinfo {author} {\bibfnamefont {Derek~J.}\ \bibnamefont
  {{Raine}}},\ }\href@noop {} {\emph {\bibinfo {title} {{Accretion Power in
  Astrophysics: Third Edition}}}}\ (\bibinfo {year} {2002})\BibitemShut
  {NoStop}%
\bibitem [{\citenamefont {Rees}\ \emph {et~al.}(1982)\citenamefont {Rees},
  \citenamefont {Phinney}, \citenamefont {Begelman},\ and\ \citenamefont
  {Blandford}}]{Rees:1982pe}%
  \BibitemOpen
  \bibfield  {author} {\bibinfo {author} {\bibfnamefont {M.~J.}\ \bibnamefont
  {Rees}}, \bibinfo {author} {\bibfnamefont {E.~S.}\ \bibnamefont {Phinney}},
  \bibinfo {author} {\bibfnamefont {M.~C.}\ \bibnamefont {Begelman}}, \ and\
  \bibinfo {author} {\bibfnamefont {R.~D.}\ \bibnamefont {Blandford}},\
  }\bibfield  {title} {\enquote {\bibinfo {title} {{Ion supported tori and the
  origin of radio jets}},}\ }\href {\doibase 10.1038/295017a0} {\bibfield
  {journal} {\bibinfo  {journal} {Nature}\ }\textbf {\bibinfo {volume} {295}},\
  \bibinfo {pages} {17--21} (\bibinfo {year} {1982})}\BibitemShut {NoStop}%
\bibitem [{\citenamefont {{Fanton}}\ \emph {et~al.}(1997)\citenamefont
  {{Fanton}}, \citenamefont {{Calvani}}, \citenamefont {{de Felice}},\ and\
  \citenamefont {{Cadez}}}]{1997PASJ...49..159F}%
  \BibitemOpen
  \bibfield  {author} {\bibinfo {author} {\bibfnamefont {Claudio}\ \bibnamefont
  {{Fanton}}}, \bibinfo {author} {\bibfnamefont {Massimo}\ \bibnamefont
  {{Calvani}}}, \bibinfo {author} {\bibfnamefont {Fernando}\ \bibnamefont {{de
  Felice}}}, \ and\ \bibinfo {author} {\bibfnamefont {Andrej}\ \bibnamefont
  {{Cadez}}},\ }\bibfield  {title} {\enquote {\bibinfo {title} {{Detecting
  Accretion Disks in Active Galactic Nuclei}},}\ }\href {\doibase
  10.1093/pasj/49.2.159} {\bibfield  {journal} {\bibinfo  {journal} {pasj}\
  }\textbf {\bibinfo {volume} {49}},\ \bibinfo {pages} {159--169} (\bibinfo
  {year} {1997})}\BibitemShut {NoStop}%
\bibitem [{\citenamefont {Marscher}\ \emph {et~al.}(2002)\citenamefont
  {Marscher}, \citenamefont {Jorstad}, \citenamefont {G{\'o}mez}, \citenamefont
  {Aller}, \citenamefont {Ter{\"a}sranta}, \citenamefont {Lister},\ and\
  \citenamefont {Stirling}}]{Marscher2002ObservationalEF}%
  \BibitemOpen
  \bibfield  {author} {\bibinfo {author} {\bibfnamefont {Alan~P.}\ \bibnamefont
  {Marscher}}, \bibinfo {author} {\bibfnamefont {Svetlana~G.}\ \bibnamefont
  {Jorstad}}, \bibinfo {author} {\bibfnamefont {Jos{\'e}~L.}\ \bibnamefont
  {G{\'o}mez}}, \bibinfo {author} {\bibfnamefont {Margo~F.}\ \bibnamefont
  {Aller}}, \bibinfo {author} {\bibfnamefont {Harri}\ \bibnamefont
  {Ter{\"a}sranta}}, \bibinfo {author} {\bibfnamefont {Matthew~L.}\
  \bibnamefont {Lister}}, \ and\ \bibinfo {author} {\bibfnamefont
  {Alastair~M.}\ \bibnamefont {Stirling}},\ }\bibfield  {title} {\enquote
  {\bibinfo {title} {Observational evidence for the accretion-disk origin for a
  radio jet in an active galaxy},}\ }\href@noop {} {\bibfield  {journal}
  {\bibinfo  {journal} {Nature}\ }\textbf {\bibinfo {volume} {417}},\ \bibinfo
  {pages} {625--627} (\bibinfo {year} {2002})}\BibitemShut {NoStop}%
\bibitem [{\citenamefont {Rezzolla}\ \emph {et~al.}(2010)\citenamefont
  {Rezzolla}, \citenamefont {Baiotti}, \citenamefont {Giacomazzo},
  \citenamefont {Link},\ and\ \citenamefont {Font}}]{Rezzolla:2010fd}%
  \BibitemOpen
  \bibfield  {author} {\bibinfo {author} {\bibfnamefont {Luciano}\ \bibnamefont
  {Rezzolla}}, \bibinfo {author} {\bibfnamefont {Luca}\ \bibnamefont
  {Baiotti}}, \bibinfo {author} {\bibfnamefont {Bruno}\ \bibnamefont
  {Giacomazzo}}, \bibinfo {author} {\bibfnamefont {David}\ \bibnamefont
  {Link}}, \ and\ \bibinfo {author} {\bibfnamefont {Jose~A.}\ \bibnamefont
  {Font}},\ }\bibfield  {title} {\enquote {\bibinfo {title} {{Accurate
  evolutions of unequal-mass neutron-star binaries: properties of the torus and
  short GRB engines}},}\ }\href {\doibase 10.1088/0264-9381/27/11/114105}
  {\bibfield  {journal} {\bibinfo  {journal} {Class. Quant. Grav.}\ }\textbf
  {\bibinfo {volume} {27}},\ \bibinfo {pages} {114105} (\bibinfo {year}
  {2010})},\ \Eprint {http://arxiv.org/abs/1001.3074} {arXiv:1001.3074 [gr-qc]}
  \BibitemShut {NoStop}%
\bibitem [{\citenamefont {Bildsten}\ \emph {et~al.}(1997)\citenamefont
  {Bildsten} \emph {et~al.}}]{Bildsten:1997vw}%
  \BibitemOpen
  \bibfield  {author} {\bibinfo {author} {\bibfnamefont {Lars}\ \bibnamefont
  {Bildsten}} \emph {et~al.},\ }\bibfield  {title} {\enquote {\bibinfo {title}
  {{Observations of accreting pulsars}},}\ }\href {\doibase 10.1086/313060}
  {\bibfield  {journal} {\bibinfo  {journal} {Astrophys. J. Suppl.}\ }\textbf
  {\bibinfo {volume} {113}},\ \bibinfo {pages} {367} (\bibinfo {year}
  {1997})},\ \Eprint {http://arxiv.org/abs/astro-ph/9707125}
  {arXiv:astro-ph/9707125} \BibitemShut {NoStop}%
\bibitem [{\citenamefont {Popham}\ and\ \citenamefont
  {Sunyaev}(2001)}]{Popham:2000qa}%
  \BibitemOpen
  \bibfield  {author} {\bibinfo {author} {\bibfnamefont {Robert}\ \bibnamefont
  {Popham}}\ and\ \bibinfo {author} {\bibfnamefont {Rashid}\ \bibnamefont
  {Sunyaev}},\ }\bibfield  {title} {\enquote {\bibinfo {title} {{Accretion disk
  boundary layers around neutron stars: X-ray production in low-mass x-ray
  binaries}},}\ }\href {\doibase 10.1086/318336} {\bibfield  {journal}
  {\bibinfo  {journal} {Astrophys. J.}\ }\textbf {\bibinfo {volume} {547}},\
  \bibinfo {pages} {355--383} (\bibinfo {year} {2001})},\ \Eprint
  {http://arxiv.org/abs/astro-ph/0004017} {arXiv:astro-ph/0004017} \BibitemShut
  {NoStop}%
\bibitem [{\citenamefont {{Rosenberg}}\ \emph {et~al.}(1975)\citenamefont
  {{Rosenberg}}, \citenamefont {{Eyles}}, \citenamefont {{Skinner}},\ and\
  \citenamefont {{Willmore}}}]{1975Natur.256..628R}%
  \BibitemOpen
  \bibfield  {author} {\bibinfo {author} {\bibfnamefont {F.~D.}\ \bibnamefont
  {{Rosenberg}}}, \bibinfo {author} {\bibfnamefont {C.~J.}\ \bibnamefont
  {{Eyles}}}, \bibinfo {author} {\bibfnamefont {G.~K.}\ \bibnamefont
  {{Skinner}}}, \ and\ \bibinfo {author} {\bibfnamefont {A.~P.}\ \bibnamefont
  {{Willmore}}},\ }\bibfield  {title} {\enquote {\bibinfo {title}
  {{Observations of a transient X-ray source with a period of 104 S}},}\ }\href
  {\doibase 10.1038/256628a0} {\bibfield  {journal} {\bibinfo  {journal}
  {\nat}\ }\textbf {\bibinfo {volume} {256}},\ \bibinfo {pages} {628--630}
  (\bibinfo {year} {1975})}\BibitemShut {NoStop}%
\bibitem [{\citenamefont {{Abramowicz}}\ and\ \citenamefont
  {{Fragile}}(2013)}]{AbramowiczFragile2013}%
  \BibitemOpen
  \bibfield  {author} {\bibinfo {author} {\bibfnamefont {M.~A.}\ \bibnamefont
  {{Abramowicz}}}\ and\ \bibinfo {author} {\bibfnamefont {P.~C.}\ \bibnamefont
  {{Fragile}}},\ }\bibfield  {title} {\enquote {\bibinfo {title} {{Foundations
  of Black Hole Accretion Disk Theory}},}\ }\href {\doibase
  10.12942/lrr-2013-1} {\bibfield  {journal} {\bibinfo  {journal} {Living
  Reviews in Relativity}\ }\textbf {\bibinfo {volume} {16}},\ \bibinfo {eid}
  {1} (\bibinfo {year} {2013})},\ \Eprint {http://arxiv.org/abs/1104.5499}
  {arXiv:1104.5499 [astro-ph.HE]} \BibitemShut {NoStop}%
\bibitem [{\citenamefont {{Rezzolla}}\ and\ \citenamefont
  {{Zanotti}}(2013)}]{Rezzolla_book:2013}%
  \BibitemOpen
  \bibfield  {author} {\bibinfo {author} {\bibfnamefont {L.}~\bibnamefont
  {{Rezzolla}}}\ and\ \bibinfo {author} {\bibfnamefont {O.}~\bibnamefont
  {{Zanotti}}},\ }\href {\doibase 10.1093/acprof:oso/9780198528906.001.0001}
  {\emph {\bibinfo {title} {Relativistic Hydrodynamics}}}\ (\bibinfo
  {publisher} {Oxford University Press},\ \bibinfo {address} {Oxford, UK},\
  \bibinfo {year} {2013})\BibitemShut {NoStop}%
\bibitem [{\citenamefont {{Fishbone}}\ and\ \citenamefont
  {{Moncrief}}(1976)}]{1976ApJ...207..962F}%
  \BibitemOpen
  \bibfield  {author} {\bibinfo {author} {\bibfnamefont {L.~G.}\ \bibnamefont
  {{Fishbone}}}\ and\ \bibinfo {author} {\bibfnamefont {V.}~\bibnamefont
  {{Moncrief}}},\ }\bibfield  {title} {\enquote {\bibinfo {title}
  {{Relativistic fluid disks in orbit around Kerr black holes.}}}\ }\href
  {\doibase 10.1086/154565} {\bibfield  {journal} {\bibinfo  {journal} {apj}\
  }\textbf {\bibinfo {volume} {207}},\ \bibinfo {pages} {962--976} (\bibinfo
  {year} {1976})}\BibitemShut {NoStop}%
\bibitem [{\citenamefont {{Abramowicz}}\ \emph {et~al.}(1978)\citenamefont
  {{Abramowicz}}, \citenamefont {{Jaroszynski}},\ and\ \citenamefont
  {{Sikora}}}]{AbramowiczJaroszynskiSikora78}%
  \BibitemOpen
  \bibfield  {author} {\bibinfo {author} {\bibfnamefont {M.}~\bibnamefont
  {{Abramowicz}}}, \bibinfo {author} {\bibfnamefont {M.}~\bibnamefont
  {{Jaroszynski}}}, \ and\ \bibinfo {author} {\bibfnamefont {M.}~\bibnamefont
  {{Sikora}}},\ }\bibfield  {title} {\enquote {\bibinfo {title} {{Relativistic,
  accreting disks}},}\ }\href@noop {} {\bibfield  {journal} {\bibinfo
  {journal} {Astron. Astrophys.}\ }\textbf {\bibinfo {volume} {63}},\ \bibinfo
  {pages} {221--224} (\bibinfo {year} {1978})}\BibitemShut {NoStop}%
\bibitem [{\citenamefont {{Font}}\ and\ \citenamefont
  {{Daigne}}(2002)}]{FoDai02}%
  \BibitemOpen
  \bibfield  {author} {\bibinfo {author} {\bibfnamefont {J.~A.}\ \bibnamefont
  {{Font}}}\ and\ \bibinfo {author} {\bibfnamefont {F.}~\bibnamefont
  {{Daigne}}},\ }\bibfield  {title} {\enquote {\bibinfo {title} {{The runaway
  instability of thick discs around black holes - I. The constant angular
  momentum case}},}\ }\href {\doibase 10.1046/j.1365-8711.2002.05515.x}
  {\bibfield  {journal} {\bibinfo  {journal} {mnras}\ }\textbf {\bibinfo
  {volume} {334}},\ \bibinfo {pages} {383--400} (\bibinfo {year} {2002})},\
  \Eprint {http://arxiv.org/abs/astro-ph/0203403} {astro-ph/0203403}
  \BibitemShut {NoStop}%
\bibitem [{\citenamefont {Lei}\ \emph {et~al.}(2009)\citenamefont {Lei},
  \citenamefont {Abramowicz}, \citenamefont {Fragile}, \citenamefont {Horak},
  \citenamefont {Machida},\ and\ \citenamefont {Straub}}]{Lei:2008ui}%
  \BibitemOpen
  \bibfield  {author} {\bibinfo {author} {\bibfnamefont {Q.}~\bibnamefont
  {Lei}}, \bibinfo {author} {\bibfnamefont {M.~A.}\ \bibnamefont {Abramowicz}},
  \bibinfo {author} {\bibfnamefont {P.~C.}\ \bibnamefont {Fragile}}, \bibinfo
  {author} {\bibfnamefont {J.}~\bibnamefont {Horak}}, \bibinfo {author}
  {\bibfnamefont {M.}~\bibnamefont {Machida}}, \ and\ \bibinfo {author}
  {\bibfnamefont {O.}~\bibnamefont {Straub}},\ }\bibfield  {title} {\enquote
  {\bibinfo {title} {{The Polish doughnuts revisited I. The angular momentum
  distribution and equipressure surfaces}},}\ }\href {\doibase
  10.1051/0004-6361/200811518} {\bibfield  {journal} {\bibinfo  {journal}
  {Astron. Astrophys.}\ }\textbf {\bibinfo {volume} {498}},\ \bibinfo {pages}
  {471--477} (\bibinfo {year} {2009})},\ \Eprint
  {http://arxiv.org/abs/0812.2467} {arXiv:0812.2467 [astro-ph]} \BibitemShut
  {NoStop}%
\bibitem [{\citenamefont {{Kuc{\'a}kov{\'a}}}\ \emph
  {et~al.}(2011)\citenamefont {{Kuc{\'a}kov{\'a}}}, \citenamefont
  {{Slan{\'y}}},\ and\ \citenamefont {{Stuchl{\'{\i}}k}}}]{KucSlStu11}%
  \BibitemOpen
  \bibfield  {author} {\bibinfo {author} {\bibfnamefont {H.}~\bibnamefont
  {{Kuc{\'a}kov{\'a}}}}, \bibinfo {author} {\bibfnamefont {P.}~\bibnamefont
  {{Slan{\'y}}}}, \ and\ \bibinfo {author} {\bibfnamefont {Z.}~\bibnamefont
  {{Stuchl{\'{\i}}k}}},\ }\bibfield  {title} {\enquote {\bibinfo {title}
  {{Toroidal configurations of perfect fluid in the
  Reissner-Nordstr{\"o}m-(anti-)de Sitter spacetimes}},}\ }\href {\doibase
  10.1088/1475-7516/2011/01/033} {\bibfield  {journal} {\bibinfo  {journal}
  {jcap}\ }\textbf {\bibinfo {volume} {1}},\ \bibinfo {eid} {033} (\bibinfo
  {year} {2011})}\BibitemShut {NoStop}%
\bibitem [{\citenamefont {Shakura}\ and\ \citenamefont
  {Sunyaev}(1973)}]{ShakuraSunyaev73}%
  \BibitemOpen
  \bibfield  {author} {\bibinfo {author} {\bibfnamefont {N.I.}\ \bibnamefont
  {Shakura}}\ and\ \bibinfo {author} {\bibfnamefont {R.A.}\ \bibnamefont
  {Sunyaev}},\ }\bibfield  {title} {\enquote {\bibinfo {title} {Black holes in
  binary systems. observational appearance},}\ }\href@noop {} {\bibfield
  {journal} {\bibinfo  {journal} {Astron. Astrophys.}\ }\textbf {\bibinfo
  {volume} {24}},\ \bibinfo {pages} {337} (\bibinfo {year} {1973})}\BibitemShut
  {NoStop}%
\bibitem [{\citenamefont {Lahiri}\ and\ \citenamefont
  {L{\"a}mmerzahl}(2019)}]{Lahiri:2019mwc}%
  \BibitemOpen
  \bibfield  {author} {\bibinfo {author} {\bibfnamefont {S.}~\bibnamefont
  {Lahiri}}\ and\ \bibinfo {author} {\bibfnamefont {C.}~\bibnamefont
  {L{\"a}mmerzahl}},\ }\bibfield  {title} {\enquote {\bibinfo {title} {{A toy
  model of viscous relativistic geometrically thick disk in Schwarzschild
  geometry}},}\ }\href@noop {} {\  (\bibinfo {year} {2019})},\ \Eprint
  {http://arxiv.org/abs/1909.10381} {arXiv:1909.10381 [gr-qc]} \BibitemShut
  {NoStop}%
\bibitem [{\citenamefont {{Kov{\'a}{\v r}}}\ \emph {et~al.}(2011)\citenamefont
  {{Kov{\'a}{\v r}}}, \citenamefont {{Slan{\'y}}}, \citenamefont
  {{Stuchl{\'{\i}}k}}, \citenamefont {{Karas}}, \citenamefont {{Cremaschini}},\
  and\ \citenamefont {{Miller}}}]{Kovar11}%
  \BibitemOpen
  \bibfield  {author} {\bibinfo {author} {\bibfnamefont {J.}~\bibnamefont
  {{Kov{\'a}{\v r}}}}, \bibinfo {author} {\bibfnamefont {P.}~\bibnamefont
  {{Slan{\'y}}}}, \bibinfo {author} {\bibfnamefont {Z.}~\bibnamefont
  {{Stuchl{\'{\i}}k}}}, \bibinfo {author} {\bibfnamefont {V.}~\bibnamefont
  {{Karas}}}, \bibinfo {author} {\bibfnamefont {C.}~\bibnamefont
  {{Cremaschini}}}, \ and\ \bibinfo {author} {\bibfnamefont {J.~C.}\
  \bibnamefont {{Miller}}},\ }\bibfield  {title} {\enquote {\bibinfo {title}
  {{Role of electric charge in shaping equilibrium configurations of fluid tori
  encircling black holes}},}\ }\href {\doibase 10.1103/PhysRevD.84.084002}
  {\bibfield  {journal} {\bibinfo  {journal} {prd}\ }\textbf {\bibinfo {volume}
  {84}},\ \bibinfo {eid} {084002} (\bibinfo {year} {2011})},\ \Eprint
  {http://arxiv.org/abs/1110.4843} {arXiv:1110.4843 [astro-ph.HE]} \BibitemShut
  {NoStop}%
\bibitem [{\citenamefont {{Kov{\'a}{\v r}}}\ \emph {et~al.}(2014)\citenamefont
  {{Kov{\'a}{\v r}}}, \citenamefont {{Slan{\'y}}}, \citenamefont
  {{Cremaschini}}, \citenamefont {{Stuchl{\'{\i}}k}}, \citenamefont {{Karas}},\
  and\ \citenamefont {{Trova}}}]{KovarTr14}%
  \BibitemOpen
  \bibfield  {author} {\bibinfo {author} {\bibfnamefont {J.}~\bibnamefont
  {{Kov{\'a}{\v r}}}}, \bibinfo {author} {\bibfnamefont {P.}~\bibnamefont
  {{Slan{\'y}}}}, \bibinfo {author} {\bibfnamefont {C.}~\bibnamefont
  {{Cremaschini}}}, \bibinfo {author} {\bibfnamefont {Z.}~\bibnamefont
  {{Stuchl{\'{\i}}k}}}, \bibinfo {author} {\bibfnamefont {V.}~\bibnamefont
  {{Karas}}}, \ and\ \bibinfo {author} {\bibfnamefont {A.}~\bibnamefont
  {{Trova}}},\ }\bibfield  {title} {\enquote {\bibinfo {title} {{Electrically
  charged matter in rigid rotation around magnetized black hole}},}\ }\href
  {\doibase 10.1103/PhysRevD.90.044029} {\bibfield  {journal} {\bibinfo
  {journal} {prd}\ }\textbf {\bibinfo {volume} {90}},\ \bibinfo {eid} {044029}
  (\bibinfo {year} {2014})},\ \Eprint {http://arxiv.org/abs/1409.0418}
  {arXiv:1409.0418 [gr-qc]} \BibitemShut {NoStop}%
\bibitem [{\citenamefont {{Slan{\'y}}}\ \emph {et~al.}(2013)\citenamefont
  {{Slan{\'y}}}, \citenamefont {{Kov{\'a}{\v r}}}, \citenamefont
  {{Stuchl{\'{\i}}k}},\ and\ \citenamefont {{Karas}}}]{slany13}%
  \BibitemOpen
  \bibfield  {author} {\bibinfo {author} {\bibfnamefont {P.}~\bibnamefont
  {{Slan{\'y}}}}, \bibinfo {author} {\bibfnamefont {J.}~\bibnamefont
  {{Kov{\'a}{\v r}}}}, \bibinfo {author} {\bibfnamefont {Z.}~\bibnamefont
  {{Stuchl{\'{\i}}k}}}, \ and\ \bibinfo {author} {\bibfnamefont
  {V.}~\bibnamefont {{Karas}}},\ }\bibfield  {title} {\enquote {\bibinfo
  {title} {{Charged Tori in Spherical Gravitational and Dipolar Magnetic
  Fields}},}\ }\href {\doibase 10.1088/0067-0049/205/1/3} {\bibfield  {journal}
  {\bibinfo  {journal} {apjs}\ }\textbf {\bibinfo {volume} {205}},\ \bibinfo
  {eid} {3} (\bibinfo {year} {2013})},\ \Eprint
  {http://arxiv.org/abs/1302.2356} {arXiv:1302.2356 [astro-ph.HE]} \BibitemShut
  {NoStop}%
\bibitem [{\citenamefont {{Komissarov}}(2006)}]{Komissarov06}%
  \BibitemOpen
  \bibfield  {author} {\bibinfo {author} {\bibfnamefont {S.~S.}\ \bibnamefont
  {{Komissarov}}},\ }\bibfield  {title} {\enquote {\bibinfo {title}
  {{Magnetized tori around Kerr black holes: analytic solutions with a toroidal
  magnetic field}},}\ }\href {\doibase 10.1111/j.1365-2966.2006.10183.x}
  {\bibfield  {journal} {\bibinfo  {journal} {mnras}\ }\textbf {\bibinfo
  {volume} {368}},\ \bibinfo {pages} {993--1000} (\bibinfo {year} {2006})},\
  \Eprint {http://arxiv.org/abs/astro-ph/0601678} {astro-ph/0601678}
  \BibitemShut {NoStop}%
\bibitem [{\citenamefont {Gimeno-Soler}\ and\ \citenamefont
  {Font}(2017)}]{Gimeno-Soler:2017qmt}%
  \BibitemOpen
  \bibfield  {author} {\bibinfo {author} {\bibfnamefont {Sergio}\ \bibnamefont
  {Gimeno-Soler}}\ and\ \bibinfo {author} {\bibfnamefont {Jos\'e~A.}\
  \bibnamefont {Font}},\ }\bibfield  {title} {\enquote {\bibinfo {title}
  {{Magnetised Polish doughnuts revisited}},}\ }\href {\doibase
  10.1051/0004-6361/201730935} {\bibfield  {journal} {\bibinfo  {journal}
  {Astron. Astrophys.}\ }\textbf {\bibinfo {volume} {607}},\ \bibinfo {pages}
  {A68} (\bibinfo {year} {2017})},\ \Eprint {http://arxiv.org/abs/1707.03867}
  {arXiv:1707.03867 [gr-qc]} \BibitemShut {NoStop}%
\bibitem [{\citenamefont {Lahiri}\ \emph {et~al.}(2021)\citenamefont {Lahiri},
  \citenamefont {Gimeno-Soler}, \citenamefont {Font},\ and\ \citenamefont
  {Mej\'\i{}as}}]{Lahiri:2020sza}%
  \BibitemOpen
  \bibfield  {author} {\bibinfo {author} {\bibfnamefont {Sayantani}\
  \bibnamefont {Lahiri}}, \bibinfo {author} {\bibfnamefont {Sergio}\
  \bibnamefont {Gimeno-Soler}}, \bibinfo {author} {\bibfnamefont {Jos\'e~A.}\
  \bibnamefont {Font}}, \ and\ \bibinfo {author} {\bibfnamefont
  {Alejandro~Mus}\ \bibnamefont {Mej\'\i{}as}},\ }\bibfield  {title} {\enquote
  {\bibinfo {title} {{Stationary models of magnetized viscous tori around a
  Schwarzschild black hole}},}\ }\href {\doibase 10.1103/PhysRevD.103.044034}
  {\bibfield  {journal} {\bibinfo  {journal} {Phys. Rev. D}\ }\textbf {\bibinfo
  {volume} {103}},\ \bibinfo {pages} {044034} (\bibinfo {year} {2021})},\
  \Eprint {http://arxiv.org/abs/2012.06835} {arXiv:2012.06835 [gr-qc]}
  \BibitemShut {NoStop}%
\bibitem [{\citenamefont {Rezzolla}\ \emph {et~al.}(2003)\citenamefont
  {Rezzolla}, \citenamefont {Zanotti},\ and\ \citenamefont
  {Font}}]{Rezzolla:2003re}%
  \BibitemOpen
  \bibfield  {author} {\bibinfo {author} {\bibfnamefont {Luciano}\ \bibnamefont
  {Rezzolla}}, \bibinfo {author} {\bibfnamefont {Olindo}\ \bibnamefont
  {Zanotti}}, \ and\ \bibinfo {author} {\bibfnamefont {Jose~A.}\ \bibnamefont
  {Font}},\ }\bibfield  {title} {\enquote {\bibinfo {title} {{Dynamics of thick
  discs around Schwarzschild-de Sitter black holes}},}\ }\href {\doibase
  10.1051/0004-6361:20031457} {\bibfield  {journal} {\bibinfo  {journal}
  {Astron. Astrophys.}\ }\textbf {\bibinfo {volume} {412}},\ \bibinfo {pages}
  {603} (\bibinfo {year} {2003})},\ \Eprint
  {http://arxiv.org/abs/gr-qc/0310045} {arXiv:gr-qc/0310045} \BibitemShut
  {NoStop}%
\bibitem [{\citenamefont {{Jefremov}}\ and\ \citenamefont
  {{Perlick}}(2016)}]{Jefremov:2016dpi}%
  \BibitemOpen
  \bibfield  {author} {\bibinfo {author} {\bibfnamefont {P.~I.}\ \bibnamefont
  {{Jefremov}}}\ and\ \bibinfo {author} {\bibfnamefont {V.}~\bibnamefont
  {{Perlick}}},\ }\bibfield  {title} {\enquote {\bibinfo {title} {{Circular
  motion in NUT space-time}},}\ }\href {\doibase
  10.1088/0264-9381/33/24/245014} {\bibfield  {journal} {\bibinfo  {journal}
  {Classical and Quantum Gravity}\ }\textbf {\bibinfo {volume} {33}},\ \bibinfo
  {eid} {245014} (\bibinfo {year} {2016})},\ \Eprint
  {http://arxiv.org/abs/1608.06218} {arXiv:1608.06218 [gr-qc]} \BibitemShut
  {NoStop}%
\bibitem [{\citenamefont {Cruz-Osorio}\ \emph {et~al.}(2021)\citenamefont
  {Cruz-Osorio}, \citenamefont {Gimeno-Soler}, \citenamefont {Font},
  \citenamefont {De~Laurentis},\ and\ \citenamefont
  {Mendoza}}]{Cruz-Osorio:2021gnz}%
  \BibitemOpen
  \bibfield  {author} {\bibinfo {author} {\bibfnamefont {Alejandro}\
  \bibnamefont {Cruz-Osorio}}, \bibinfo {author} {\bibfnamefont {Sergio}\
  \bibnamefont {Gimeno-Soler}}, \bibinfo {author} {\bibfnamefont {Jos\'e~A.}\
  \bibnamefont {Font}}, \bibinfo {author} {\bibfnamefont {Mariafelicia}\
  \bibnamefont {De~Laurentis}}, \ and\ \bibinfo {author} {\bibfnamefont
  {Sergio}\ \bibnamefont {Mendoza}},\ }\bibfield  {title} {\enquote {\bibinfo
  {title} {{Magnetized discs and photon rings around Yukawa-like black
  holes}},}\ }\href {\doibase 10.1103/PhysRevD.103.124009} {\bibfield
  {journal} {\bibinfo  {journal} {Phys. Rev. D}\ }\textbf {\bibinfo {volume}
  {103}},\ \bibinfo {pages} {124009} (\bibinfo {year} {2021})},\ \Eprint
  {http://arxiv.org/abs/2102.10150} {arXiv:2102.10150 [astro-ph.HE]}
  \BibitemShut {NoStop}%
\bibitem [{\citenamefont {Wielgus}\ \emph {et~al.}(2022)\citenamefont
  {Wielgus}, \citenamefont {Lancova}, \citenamefont {Straub}, \citenamefont
  {Kluzniak}, \citenamefont {Narayan}, \citenamefont {Abarca}, \citenamefont
  {Rozanska}, \citenamefont {Vincent}, \citenamefont {Torok},\ and\
  \citenamefont {Abramowicz}}]{Wielgus:2022qij}%
  \BibitemOpen
  \bibfield  {author} {\bibinfo {author} {\bibfnamefont {Maciek}\ \bibnamefont
  {Wielgus}}, \bibinfo {author} {\bibfnamefont {Debora}\ \bibnamefont
  {Lancova}}, \bibinfo {author} {\bibfnamefont {Odele}\ \bibnamefont {Straub}},
  \bibinfo {author} {\bibfnamefont {Wlodek}\ \bibnamefont {Kluzniak}}, \bibinfo
  {author} {\bibfnamefont {Ramesh}\ \bibnamefont {Narayan}}, \bibinfo {author}
  {\bibfnamefont {David}\ \bibnamefont {Abarca}}, \bibinfo {author}
  {\bibfnamefont {Agata}\ \bibnamefont {Rozanska}}, \bibinfo {author}
  {\bibfnamefont {Frederic}\ \bibnamefont {Vincent}}, \bibinfo {author}
  {\bibfnamefont {Gabriel}\ \bibnamefont {Torok}}, \ and\ \bibinfo {author}
  {\bibfnamefont {Marek}\ \bibnamefont {Abramowicz}},\ }\bibfield  {title}
  {\enquote {\bibinfo {title} {{Observational properties of puffy discs:
  radiative GRMHD spectra of mildly sub-Eddington accretion}},}\ }\href
  {\doibase 10.1093/mnras/stac1317} {\bibfield  {journal} {\bibinfo  {journal}
  {Mon. Not. Roy. Astron. Soc.}\ }\textbf {\bibinfo {volume} {514}},\ \bibinfo
  {pages} {780--789} (\bibinfo {year} {2022})},\ \Eprint
  {http://arxiv.org/abs/2202.08831} {arXiv:2202.08831 [astro-ph.HE]}
  \BibitemShut {NoStop}%
\bibitem [{\citenamefont {Bahamonde}\ \emph {et~al.}(2022)\citenamefont
  {Bahamonde}, \citenamefont {Faraji}, \citenamefont {Hackmann},\ and\
  \citenamefont {Pfeifer}}]{Bahamonde:2022jue}%
  \BibitemOpen
  \bibfield  {author} {\bibinfo {author} {\bibfnamefont {Sebastian}\
  \bibnamefont {Bahamonde}}, \bibinfo {author} {\bibfnamefont {Shokoufe}\
  \bibnamefont {Faraji}}, \bibinfo {author} {\bibfnamefont {Eva}\ \bibnamefont
  {Hackmann}}, \ and\ \bibinfo {author} {\bibfnamefont {Christian}\
  \bibnamefont {Pfeifer}},\ }\bibfield  {title} {\enquote {\bibinfo {title}
  {{Thick accretion disk configurations in the Born-Infeld teleparallel
  gravity}},}\ }\href {\doibase 10.1103/PhysRevD.106.084046} {\bibfield
  {journal} {\bibinfo  {journal} {Phys. Rev. D}\ }\textbf {\bibinfo {volume}
  {106}},\ \bibinfo {pages} {084046} (\bibinfo {year} {2022})},\ \Eprint
  {http://arxiv.org/abs/2209.00020} {arXiv:2209.00020 [gr-qc]} \BibitemShut
  {NoStop}%
\bibitem [{\citenamefont {Cassing}\ and\ \citenamefont
  {Rezzolla}(2023)}]{Cassing:2023bpt}%
  \BibitemOpen
  \bibfield  {author} {\bibinfo {author} {\bibfnamefont {Marie}\ \bibnamefont
  {Cassing}}\ and\ \bibinfo {author} {\bibfnamefont {Luciano}\ \bibnamefont
  {Rezzolla}},\ }\bibfield  {title} {\enquote {\bibinfo {title} {{Equilibrium
  non-selfgravitating tori around black holes in parameterised spherically
  symmetric spacetimes}},}\ }\href {\doibase 10.1093/mnras/stad1039} {\
  (\bibinfo {year} {2023}),\ 10.1093/mnras/stad1039},\ \Eprint
  {http://arxiv.org/abs/2302.09135} {arXiv:2302.09135 [gr-qc]} \BibitemShut
  {NoStop}%
\bibitem [{\citenamefont {{Abramowicz}}\ \emph {et~al.}(1984)\citenamefont
  {{Abramowicz}}, \citenamefont {{Curir}}, \citenamefont
  {{Schwarzenberg-Czerny}},\ and\ \citenamefont
  {{Wilson}}}]{1984MNRAS.208..279A}%
  \BibitemOpen
  \bibfield  {author} {\bibinfo {author} {\bibfnamefont {M.~A.}\ \bibnamefont
  {{Abramowicz}}}, \bibinfo {author} {\bibfnamefont {A.}~\bibnamefont
  {{Curir}}}, \bibinfo {author} {\bibfnamefont {A.}~\bibnamefont
  {{Schwarzenberg-Czerny}}}, \ and\ \bibinfo {author} {\bibfnamefont {R.~E.}\
  \bibnamefont {{Wilson}}},\ }\bibfield  {title} {\enquote {\bibinfo {title}
  {{Self-gravity and the global structure of accretion discs}},}\ }\href
  {\doibase 10.1093/mnras/208.2.279} {\bibfield  {journal} {\bibinfo  {journal}
  {mnras}\ }\textbf {\bibinfo {volume} {208}},\ \bibinfo {pages} {279--291}
  (\bibinfo {year} {1984})}\BibitemShut {NoStop}%
\bibitem [{\citenamefont {Lodato}(2007)}]{Lodato:2007hmf}%
  \BibitemOpen
  \bibfield  {author} {\bibinfo {author} {\bibfnamefont {G.}~\bibnamefont
  {Lodato}},\ }\bibfield  {title} {\enquote {\bibinfo {title}
  {{Self-gravitating accretion discs}},}\ }\href {\doibase
  10.1393/ncr/i2007-10022-x} {\bibfield  {journal} {\bibinfo  {journal} {Riv.
  Nuovo Cim.}\ }\textbf {\bibinfo {volume} {30}},\ \bibinfo {pages} {293--353}
  (\bibinfo {year} {2007})},\ \Eprint {http://arxiv.org/abs/0801.3848}
  {arXiv:0801.3848 [astro-ph]} \BibitemShut {NoStop}%
\bibitem [{\citenamefont {Stergioulas}(2011)}]{Stergioulas:2011ga}%
  \BibitemOpen
  \bibfield  {author} {\bibinfo {author} {\bibfnamefont {Nikolaos}\
  \bibnamefont {Stergioulas}},\ }\bibfield  {title} {\enquote {\bibinfo {title}
  {{An Improved method for constructing models of self-gravitating tori around
  black holes}},}\ }\href {\doibase 10.1142/S021827181101944X} {\bibfield
  {journal} {\bibinfo  {journal} {Int. J. Mod. Phys. D}\ }\textbf {\bibinfo
  {volume} {20}},\ \bibinfo {pages} {1251--1263} (\bibinfo {year} {2011})},\
  \Eprint {http://arxiv.org/abs/1104.3685} {arXiv:1104.3685 [gr-qc]}
  \BibitemShut {NoStop}%
\bibitem [{\citenamefont {Mach}(2007)}]{Mach:2007zz}%
  \BibitemOpen
  \bibfield  {author} {\bibinfo {author} {\bibfnamefont {Patryk}\ \bibnamefont
  {Mach}},\ }\bibfield  {title} {\enquote {\bibinfo {title} {{Selfgravitation
  and stability in spherical accretion}},}\ }\href@noop {} {\bibfield
  {journal} {\bibinfo  {journal} {Acta Phys. Polon. B}\ }\textbf {\bibinfo
  {volume} {38}},\ \bibinfo {pages} {3935--3945} (\bibinfo {year}
  {2007})}\BibitemShut {NoStop}%
\bibitem [{\citenamefont {Mach}\ and\ \citenamefont
  {Malec}(2022)}]{Mach:2022wrk}%
  \BibitemOpen
  \bibfield  {author} {\bibinfo {author} {\bibfnamefont {Patryk}\ \bibnamefont
  {Mach}}\ and\ \bibinfo {author} {\bibfnamefont {Edward}\ \bibnamefont
  {Malec}},\ }\bibfield  {title} {\enquote {\bibinfo {title} {{Steady critical
  accretion onto black holes: Self-gravity and sonic point characteristics}},}\
  }\href {\doibase 10.1103/PhysRevD.105.104012} {\bibfield  {journal} {\bibinfo
   {journal} {Phys. Rev. D}\ }\textbf {\bibinfo {volume} {105}},\ \bibinfo
  {pages} {104012} (\bibinfo {year} {2022})},\ \Eprint
  {http://arxiv.org/abs/2202.01524} {arXiv:2202.01524 [gr-qc]} \BibitemShut
  {NoStop}%
\bibitem [{\citenamefont {{Halbwachs}}(1960)}]{1960PThPh..24..291H}%
  \BibitemOpen
  \bibfield  {author} {\bibinfo {author} {\bibfnamefont {F.}~\bibnamefont
  {{Halbwachs}}},\ }\bibfield  {title} {\enquote {\bibinfo {title} {{Lagrangian
  Formalism for a Classical Relativistic Particle Endowed with Internal
  Structure}},}\ }\href {\doibase 10.1143/PTP.24.291} {\bibfield  {journal}
  {\bibinfo  {journal} {Progress of Theoretical Physics}\ }\textbf {\bibinfo
  {volume} {24}},\ \bibinfo {pages} {291--307} (\bibinfo {year}
  {1960})}\BibitemShut {NoStop}%
\bibitem [{\citenamefont {Weyssenhoff}\ and\ \citenamefont
  {Raabe}(1947)}]{Weyssenhoff:1947iua}%
  \BibitemOpen
  \bibfield  {author} {\bibinfo {author} {\bibfnamefont {Jan}\ \bibnamefont
  {Weyssenhoff}}\ and\ \bibinfo {author} {\bibfnamefont {A.}~\bibnamefont
  {Raabe}},\ }\bibfield  {title} {\enquote {\bibinfo {title} {{Relativistic
  dynamics of spin-fluids and spin-particles}},}\ }\href@noop {} {\bibfield
  {journal} {\bibinfo  {journal} {Acta Phys. Polon.}\ }\textbf {\bibinfo
  {volume} {9}},\ \bibinfo {pages} {7--18} (\bibinfo {year}
  {1947})}\BibitemShut {NoStop}%
\bibitem [{\citenamefont {Obukhov}\ and\ \citenamefont
  {Korotkii}(1987)}]{Obukhov:1987yu}%
  \BibitemOpen
  \bibfield  {author} {\bibinfo {author} {\bibfnamefont {Y.~N.}\ \bibnamefont
  {Obukhov}}\ and\ \bibinfo {author} {\bibfnamefont {V.~A.}\ \bibnamefont
  {Korotkii}},\ }\bibfield  {title} {\enquote {\bibinfo {title} {{The
  Weyssenhoff fluid in Einstein-Cartan theory}},}\ }\href {\doibase
  10.1088/0264-9381/4/6/021} {\bibfield  {journal} {\bibinfo  {journal} {Class.
  Quant. Grav.}\ }\textbf {\bibinfo {volume} {4}},\ \bibinfo {pages}
  {1633--1657} (\bibinfo {year} {1987})}\BibitemShut {NoStop}%
\bibitem [{\citenamefont {Obukhov}\ and\ \citenamefont
  {Piskareva}(1989)}]{ObukhovPiskareva1989}%
  \BibitemOpen
  \bibfield  {author} {\bibinfo {author} {\bibfnamefont {Y.N.}\ \bibnamefont
  {Obukhov}}\ and\ \bibinfo {author} {\bibfnamefont {O.B.}\ \bibnamefont
  {Piskareva}},\ }\bibfield  {title} {\enquote {\bibinfo {title} {Spinning
  fluid in general relativity},}\ }\href {\doibase 10.1088/0264-9381/6/2/002}
  {\bibfield  {journal} {\bibinfo  {journal} {Classical and Quantum Gravity}\
  }\textbf {\bibinfo {volume} {6}},\ \bibinfo {pages} {L15--L19} (\bibinfo
  {year} {1989})}\BibitemShut {NoStop}%
\bibitem [{\citenamefont {Mathisson}(1937)}]{Mathisson:1937zz}%
  \BibitemOpen
  \bibfield  {author} {\bibinfo {author} {\bibfnamefont {Myron}\ \bibnamefont
  {Mathisson}},\ }\bibfield  {title} {\enquote {\bibinfo {title} {{Neue
  mechanik materieller systemes}},}\ }\href@noop {} {\bibfield  {journal}
  {\bibinfo  {journal} {Acta Phys. Polon.}\ }\textbf {\bibinfo {volume} {6}},\
  \bibinfo {pages} {163--2900} (\bibinfo {year} {1937})}\BibitemShut {NoStop}%
\bibitem [{\citenamefont {{Papapetrou}}(1951)}]{1951RSPSA.209..248P}%
  \BibitemOpen
  \bibfield  {author} {\bibinfo {author} {\bibfnamefont {A.}~\bibnamefont
  {{Papapetrou}}},\ }\bibfield  {title} {\enquote {\bibinfo {title} {{Spinning
  Test-Particles in General Relativity. I}},}\ }\href {\doibase
  10.1098/rspa.1951.0200} {\bibfield  {journal} {\bibinfo  {journal}
  {Proceedings of the Royal Society of London Series A}\ }\textbf {\bibinfo
  {volume} {209}},\ \bibinfo {pages} {248--258} (\bibinfo {year}
  {1951})}\BibitemShut {NoStop}%
\bibitem [{\citenamefont {{Dixon}}(1970)}]{1970RSPSA.314..499D}%
  \BibitemOpen
  \bibfield  {author} {\bibinfo {author} {\bibfnamefont {W.~G.}\ \bibnamefont
  {{Dixon}}},\ }\bibfield  {title} {\enquote {\bibinfo {title} {{Dynamics of
  Extended Bodies in General Relativity. I. Momentum and Angular Momentum}},}\
  }\href {\doibase 10.1098/rspa.1970.0020} {\bibfield  {journal} {\bibinfo
  {journal} {Proceedings of the Royal Society of London Series A}\ }\textbf
  {\bibinfo {volume} {314}},\ \bibinfo {pages} {499--527} (\bibinfo {year}
  {1970})}\BibitemShut {NoStop}%
\bibitem [{\citenamefont {{Pirani}}(1956)}]{1956AcPP...15..389P}%
  \BibitemOpen
  \bibfield  {author} {\bibinfo {author} {\bibfnamefont {F.~A.~E.}\
  \bibnamefont {{Pirani}}},\ }\bibfield  {title} {\enquote {\bibinfo {title}
  {{On the physical significance of the Riemann tensor}},}\ }\href@noop {}
  {\bibfield  {journal} {\bibinfo  {journal} {Acta Physica Polonica}\ }\textbf
  {\bibinfo {volume} {15}},\ \bibinfo {pages} {389--405} (\bibinfo {year}
  {1956})}\BibitemShut {NoStop}%
\bibitem [{\citenamefont {Frenkel}(1926)}]{Frenkel:1926zz}%
  \BibitemOpen
  \bibfield  {author} {\bibinfo {author} {\bibfnamefont {J.}~\bibnamefont
  {Frenkel}},\ }\bibfield  {title} {\enquote {\bibinfo {title} {{Die
  Elektrodynamik des rotierenden Elektrons}},}\ }\href {\doibase
  10.1007/BF01397099} {\bibfield  {journal} {\bibinfo  {journal} {Z. Phys.}\
  }\textbf {\bibinfo {volume} {37}},\ \bibinfo {pages} {243--262} (\bibinfo
  {year} {1926})}\BibitemShut {NoStop}%
\bibitem [{\citenamefont {Tulczyjew}(1959)}]{Tulczyjew}%
  \BibitemOpen
  \bibfield  {author} {\bibinfo {author} {\bibfnamefont {W.~M.}\ \bibnamefont
  {Tulczyjew}},\ }\bibfield  {title} {\enquote {\bibinfo {title} {{Motion of
  multipole particles in general relativity theory binaries}},}\ }\href@noop {}
  {\bibfield  {journal} {\bibinfo  {journal} {Acta Phys. Pol.}\ }\textbf
  {\bibinfo {volume} {18}},\ \bibinfo {pages} {393--409} (\bibinfo {year}
  {1959})}\BibitemShut {NoStop}%
\bibitem [{\citenamefont {Dixon}(1970)}]{Dixon:1970zza}%
  \BibitemOpen
  \bibfield  {author} {\bibinfo {author} {\bibfnamefont {W.~G.}\ \bibnamefont
  {Dixon}},\ }\bibfield  {title} {\enquote {\bibinfo {title} {{Dynamics of
  extended bodies in general relativity. I. Momentum and angular momentum}},}\
  }\href {\doibase 10.1098/rspa.1970.0020} {\bibfield  {journal} {\bibinfo
  {journal} {Proc. Roy. Soc. Lond. A}\ }\textbf {\bibinfo {volume} {314}},\
  \bibinfo {pages} {499--527} (\bibinfo {year} {1970})}\BibitemShut {NoStop}%
\bibitem [{\citenamefont {Newton}\ and\ \citenamefont
  {Wigner}(1949)}]{newton1949localized}%
  \BibitemOpen
  \bibfield  {author} {\bibinfo {author} {\bibfnamefont {Theodore~Duddell}\
  \bibnamefont {Newton}}\ and\ \bibinfo {author} {\bibfnamefont {Eugene~P}\
  \bibnamefont {Wigner}},\ }\bibfield  {title} {\enquote {\bibinfo {title}
  {Localized states for elementary systems},}\ }\href@noop {} {\bibfield
  {journal} {\bibinfo  {journal} {Reviews of Modern Physics}\ }\textbf
  {\bibinfo {volume} {21}},\ \bibinfo {pages} {400} (\bibinfo {year}
  {1949})}\BibitemShut {NoStop}%
\bibitem [{\citenamefont {Kyrian}\ and\ \citenamefont
  {Semerak}(2007)}]{Kyrian:2007zz}%
  \BibitemOpen
  \bibfield  {author} {\bibinfo {author} {\bibfnamefont {K}~\bibnamefont
  {Kyrian}}\ and\ \bibinfo {author} {\bibfnamefont {O}~\bibnamefont
  {Semerak}},\ }\bibfield  {title} {\enquote {\bibinfo {title} {{Spinning test
  particles in a Kerr field}},}\ }\href {\doibase
  10.1111/j.1365-2966.2007.12502.x} {\bibfield  {journal} {\bibinfo  {journal}
  {Mon. Not. Roy. Astron. Soc.}\ }\textbf {\bibinfo {volume} {382}},\ \bibinfo
  {pages} {1922} (\bibinfo {year} {2007})}\BibitemShut {NoStop}%
\bibitem [{\citenamefont {{Rezzolla}}\ \emph {et~al.}(2003)\citenamefont
  {{Rezzolla}}, \citenamefont {{Zanotti}},\ and\ \citenamefont
  {{Font}}}]{RezZanFon03}%
  \BibitemOpen
  \bibfield  {author} {\bibinfo {author} {\bibfnamefont {L.}~\bibnamefont
  {{Rezzolla}}}, \bibinfo {author} {\bibfnamefont {O.}~\bibnamefont
  {{Zanotti}}}, \ and\ \bibinfo {author} {\bibfnamefont {J.~A.}\ \bibnamefont
  {{Font}}},\ }\bibfield  {title} {\enquote {\bibinfo {title} {{Dynamics of
  thick discs around Schwarzschild-de Sitter black holes}},}\ }\href {\doibase
  10.1051/0004-6361:20031457} {\bibfield  {journal} {\bibinfo  {journal} {aap}\
  }\textbf {\bibinfo {volume} {412}},\ \bibinfo {pages} {603--613} (\bibinfo
  {year} {2003})},\ \Eprint {http://arxiv.org/abs/gr-qc/0310045}
  {gr-qc/0310045} \BibitemShut {NoStop}%
\bibitem [{\citenamefont {{Abramowicz}}\ and\ \citenamefont
  {{Calvani}}(1979)}]{1979MNRAS.189..621A}%
  \BibitemOpen
  \bibfield  {author} {\bibinfo {author} {\bibfnamefont {M.~A.}\ \bibnamefont
  {{Abramowicz}}}\ and\ \bibinfo {author} {\bibfnamefont {M.}~\bibnamefont
  {{Calvani}}},\ }\bibfield  {title} {\enquote {\bibinfo {title} {{Spinning
  particles orbiting the Kerr black hole}},}\ }\href {\doibase
  10.1093/mnras/189.3.621} {\bibfield  {journal} {\bibinfo  {journal} {mnras}\
  }\textbf {\bibinfo {volume} {189}},\ \bibinfo {pages} {621--626} (\bibinfo
  {year} {1979})}\BibitemShut {NoStop}%
\bibitem [{\citenamefont {Semerak}(1999)}]{Semerak:1999qc}%
  \BibitemOpen
  \bibfield  {author} {\bibinfo {author} {\bibfnamefont {O.}~\bibnamefont
  {Semerak}},\ }\bibfield  {title} {\enquote {\bibinfo {title} {{Spinning test
  particles in a Kerr field. 1.}}}\ }\href {\doibase
  10.1046/j.1365-8711.1999.02754.x} {\bibfield  {journal} {\bibinfo  {journal}
  {Mon. Not. Roy. Astron. Soc.}\ }\textbf {\bibinfo {volume} {308}},\ \bibinfo
  {pages} {863--875} (\bibinfo {year} {1999})}\BibitemShut {NoStop}%
\bibitem [{\citenamefont {{Garcia}}\ \emph {et~al.}(2020)\citenamefont
  {{Garcia}}, \citenamefont {{Mendoza}}, \citenamefont {{Bautista}},
  \citenamefont {{Kallman}}, \citenamefont {{Deprince}}, \citenamefont
  {{Palmeri}},\ and\ \citenamefont {{Quinet}}}]{2020AAS...23534604G}%
  \BibitemOpen
  \bibfield  {author} {\bibinfo {author} {\bibfnamefont {J.~A.}\ \bibnamefont
  {{Garcia}}}, \bibinfo {author} {\bibfnamefont {C.}~\bibnamefont {{Mendoza}}},
  \bibinfo {author} {\bibfnamefont {M.}~\bibnamefont {{Bautista}}}, \bibinfo
  {author} {\bibfnamefont {T.}~\bibnamefont {{Kallman}}}, \bibinfo {author}
  {\bibfnamefont {J.}~\bibnamefont {{Deprince}}}, \bibinfo {author}
  {\bibfnamefont {P.}~\bibnamefont {{Palmeri}}}, \ and\ \bibinfo {author}
  {\bibfnamefont {P.}~\bibnamefont {{Quinet}}},\ }\bibfield  {title} {\enquote
  {\bibinfo {title} {{High-Density Plasma Effects in Accretion Disks around
  Black Holes}},}\ }in\ \href@noop {} {\emph {\bibinfo {booktitle} {American
  Astronomical Society Meeting Abstracts \#235}}},\ \bibinfo {series} {American
  Astronomical Society Meeting Abstracts}, Vol.\ \bibinfo {volume} {235}\
  (\bibinfo {year} {2020})\ p.\ \bibinfo {pages} {346.04}\BibitemShut {NoStop}%
\bibitem [{\citenamefont
  {Lukes-Gerakopoulos}(2017)}]{Lukes-Gerakopoulos:2016udm}%
  \BibitemOpen
  \bibfield  {author} {\bibinfo {author} {\bibfnamefont {Georgios}\
  \bibnamefont {Lukes-Gerakopoulos}},\ }\bibfield  {title} {\enquote {\bibinfo
  {title} {{Spinning particles moving around black holes: integrability and
  chaos}},}\ }in\ \href {\doibase 10.1142/9789813226609_0209} {\emph {\bibinfo
  {booktitle} {{14th Marcel Grossmann Meeting on Recent Developments in
  Theoretical and Experimental General Relativity, Astrophysics, and
  Relativistic Field Theories}}}},\ Vol.~\bibinfo {volume} {2}\ (\bibinfo
  {year} {2017})\ pp.\ \bibinfo {pages} {1960--1965},\ \Eprint
  {http://arxiv.org/abs/1606.09430} {arXiv:1606.09430 [gr-qc]} \BibitemShut
  {NoStop}%
\end{thebibliography}%
\end{document}